\documentclass[twocolumn,twoside]{IEEEtran}
\usepackage{mathpple}
\usepackage{times}
\usepackage{amsmath}  % Define \boldsymbol (in amsbsy too) and align
\usepackage{amssymb}  % Define \mathbb (in amsfonts too)
\usepackage{mathrsfs} % Define \mathscr, a script font
\usepackage{theorem}  % Helps in rendering theorems, etc.
\usepackage{cite}     % Gives multiple references as intervals
\usepackage{comment}  % Comments out with \begin{comment} ... \end{comment}
\usepackage{upref}
\usepackage{amsfonts}
\usepackage{verbatim}
\usepackage[dvipsnames,usenames]{color}
\usepackage{enumerate}
\usepackage{cuted}
\usepackage[shortlabels]{enumitem}

\usepackage{algorithm}
\usepackage{times,color}
\usepackage{graphicx}
\usepackage{float}
\usepackage{color}
\usepackage{mathtools}
\usepackage{graphicx}
\usepackage{listings}
\usepackage{tikz}
\usepackage{multicol}
\usepackage{psfrag}

\usepackage{pgfplots}
\usepackage{hhline}
\usepackage{xcolor,colortbl}
\usepackage{algorithm}
\usepackage[noend]{algpseudocode}

\usepackage{wrapfig}
\usepackage{subfigure}
\usepackage[top=0.58in, bottom=0.58in, left=0.6in, right=0.6in]{geometry}
\usepackage{kantlipsum,cuted}
\usepackage[draft,ulem=normalem]{changes}
\usepackage[all]{xy}
\usepackage{algorithmicx}
\algrenewcommand\alglinenumber[1]{\scriptsize #1:}
\algrenewcommand\algorithmicindent{1em}%
\usepackage{latexsym}
\usepackage{multirow}% http://ctan.org/pkg/multirow

\usepackage{tabu}
\usepackage{lipsum}
\usepackage{blkarray, bigstrut}
\usepackage{comment}

\newcommand{\RN}[1]{%
	\textup{\uppercase\expandafter{\romannumeral#1}}%
}

\newtheorem{innercustomgeneric}{\customgenericname}
\providecommand{\customgenericname}{}
\newcommand{\newcustomtheorem}[2]{%
	\newenvironment{#1}[1]
	{%
		\renewcommand\customgenericname{#2}%
		\renewcommand\theinnercustomgeneric{##1}%
		\innercustomgeneric
	}
	{\endinnercustomgeneric}
}
\newcustomtheorem{customproposition}{Proposition}
\newcustomtheorem{customtheorem}{Theorem}
\newcustomtheorem{customlemma}{Lemma}
\newcustomtheorem{customclaim}{Claim}

\makeatletter
\def\BState{\State\hskip-\ALG@thistlm}

\makeatother

\algnewcommand{\Initialize}[1]{%
	\State \textbf{Initialize:}
	\Statex \hspace*{\algorithmicindent}\parbox[t]{.8\linewidth}{\raggedright #1}
}

\newcommand{\be}[1]{\begin{equation}\label{#1}}
\newcommand{\ee}{\end{equation}}

\newcommand{\bc}{\begin{center}}
\newcommand{\ec}{\end{center}}

\newcommand{\cC}{{\cal C}}

\newcommand{\cG}{{\cal G}}

\newcommand{\cR}{{\cal R}}
\newcommand{\cS}{{\cal S}}

\newcommand{\cU}{{\cal U}}

\renewcommand{\leq}{\leqslant}

\renewcommand{\geq}{\geqslant}

\newcommand{\set}[1]{\left\{ #1 \right\}}

\newcommand{\F}{\mathbb{F}}
\newcommand{\Fq}{\smash{\mathbb{F}_{\!q}}}

\newcommand{\Z}{\mathbb{Z}}
\newcommand{\N}{\mathbb{N}}

\newcommand{\ipn}[1]{\left\langle #1\right\rangle _n}

\newcommand{\Cref}[1]{Co\-rol\-la\-ry\,\ref{#1}}

\theoremstyle{plain} \theorembodyfont{\normalfont\slshape}

\newtheorem{thm}{Theorem$\!$}
\newenvironment{theorem}{\begin{thm}\hspace*{-1ex}{\bf.}}{\end{thm}}

\newtheorem{prop}[thm]{Proposition$\!$}

\newtheorem{lem}[thm]{Lemma$\!$}
\newenvironment{lemma}{\begin{lem}\hspace*{-1ex}{\bf.}}{\end{lem}}

\newtheorem{cor}[thm]{Corollary$\!$}
\newenvironment{corollary}{\begin{cor}\hspace*{-1ex}{\bf.}}{\end{cor}}

\newtheorem{cons}[thm]{Construction$\!$}

\newtheorem{defi}[thm]{Definition$\!$}

\newtheorem{cl}{Claim}
\newenvironment{claim}{\begin{cl}\hspace*{-1ex}{\bf .}}{\end{cl}}
\theorembodyfont{\normalfont}

\newtheorem{exam}{Example$\!$}
\newenvironment{example}{\begin{exam}\hspace*{-1ex}{\bf .}}{\end{exam}}

\newtheorem{remrk}{Remark$\!$}

\newtheorem{Construction}{Construction}

\definecolor{Codecolor}{named}{White}  %{Tan}
\newcommand{\Copen}{\mbox{\{\kern-5.50pt\{}}
\newcommand{\Cclose}{\mbox{\}\kern-5.50pt\}}}
\newcommand{\Cslash}{\mbox{$\backslash\kern-6.02pt\backslash$}}

\newcommand{\gc}{\cG\textmd{-}}
\newcommand{\uc}{\cU\textmd{-}}

\begin{document}
		
		\title{Double and Triple Node-Erasure-Correcting Codes over Graphs}
		
		\author{\large Lev~Yohananov,~\IEEEmembership{Student Member,~IEEE}, Yuval Efron,  and Eitan~Yaakobi,~\IEEEmembership{Senior Member,~IEEE} 
			\thanks{L. Yohananov, Y. Efron, and E. Yaakobi are with the Department of Computer Science, Technion --- Israel Institute of Technology, Haifa 32000, Israel (e-mail: \texttt{\{levyohananov,szxrtde,yaakobi\}@cs.technion.ac.il}).}
\vspace{-1ex}		}

		\maketitle		
\begin{abstract}
	In this paper we study array-based codes over graphs for correcting multiple node failures. These codes have applications to neural networks, associative memories, and distributed storage systems. We assume that the information is stored on the edges of a complete undirected graph and a \emph{node failure} is the event where all the edges in the neighborhood of a given node have been erased. A code over graphs is called \emph{$\rho$-node-erasure-correcting} if it allows to reconstruct the erased edges upon the failure of any $\rho$ nodes or less. We present a binary optimal construction for double-node-erasure correction together with an efficient decoding algorithm, when the number of nodes is a prime number. Furthermore, we extend this construction for triple-node-erasure-correcting codes when the number of nodes is a prime number and two is a primitive element in $\Z_n$. These codes are at most a single bit away from optimality.
\end{abstract}
\vspace{0ex}

\section{Introduction} \label{sec:intro}
Networks and distributed storage systems are usually
represented as graphs with the information stored
in the nodes (vertices) of the graph.
In our recent work~\cite{YY17,YY18,YY18b},
we have introduced a new model which assumes that
the information is stored on the \emph{edges}.
This setup is motivated by several information systems.
For example, in \emph{neural networks}, the neural units are connected
via \emph{links} which store and transmit information between
the neural units~\cite{Hopfield:1988:NNP:65669.104422}.
Similarly, in associative memories, the information is stored
by associations between different data items~\cite{6283016}.
Furthermore, representing information in a graph can model
a distributed storage system~\cite{NetworkCoding} while 
every two nodes can be connected by a link that represents the 
information that is shared by the nodes.

In~\cite{YY17,YY18,YY18b}, we introduced the notion of
\emph{codes over graphs},
which is a class of codes
storing the information on the edges of a complete 
undirected graph (including self-loops).
Thus, each codeword is a labeled graph with~$n$ \emph{nodes} (vertices)
and each of the $\binom{n+1}{2}$ edges stores
a symbol over an alphabet~$\Sigma$.
A \emph{node failure} is the event where all the edges incident
with a given node have been erased,
and a code over graphs is called
\emph{$\rho$-node-erasure-correcting} if it allows
to reconstruct the contents of the erased edges
upon the failure of any $\rho$ nodes or less.

The information stored in
a complete undirected graph can be represented by 
an $n \times n$ symmetric array and a failure of the $i$th node 
corresponds to the 
erasure of the $i$th row and $i$th column in the array. Hence, 
this problem is translated to the problem of correcting
\emph{symmetric crisscross erasures}
in square symmetric arrays~\cite{DBLP:journals/tit/Roth91}. 
By the Singleton bound, the number of
\emph{redundancy edges} (i.e., redundancy symbols in the array) 
of every $\rho$-node-erasure-correcting code
must be at least $n\rho - \binom{\rho}{2}$, and a code
meeting this bound will be referred as \emph{optimal}.
While the construction of optimal codes is easily accomplished
by MDS codes,
their alphabet size must be at least the order of $n^2$, and
the task of constructing optimal (or close to optimal) codes over graphs
over smaller alphabets remains an intriguing problem. 

A natural approach to address this problem is by using the wide
existing knowledge on array code constructions such as
\cite{DBLP:journals/tc/BlaumBBM95,ButterflyISIT,DBLP:journals/tc/HuangX08,journals/jct/Schmidt10,RDP,DBLP:conf/icc/MohammedVHC10,Raviv16,DBLP:journals/tit/Roth91,S15,DBLP:journals/tit/TamoWB13,DBLP:journals/tit/XuBBW99}.
However, the setup of codes over graphs differs from
that of classical array codes in two respects.
First, the arrays are symmetric, and, secondly, a failure of
the $i$th node in the graph corresponds to the failure of
the $i$th row and the $i$th column (for the same~$i$) in the array.
Most existing constructions of array codes are not designed
for symmetric arrays, and they do not support this special 
row--column failure model.
However, it is still possible to use existing code constructions
and modify them to the special structure of
the above erasure model in graphs, as was done in~\cite{YY17},\cite{YY18b}.
More specifically, based upon product codes~\cite{ProductCode1},\cite{ProductCode2}, 
a construction of optimal codes whose alphabet size grows only
linearly with~$n$ has been proposed. Additionally, using rank-metric codes~\cite{DBLP:journals/tit/Roth91,journals/jct/Schmidt10,S15}, binary codes over graphs were designed, 
however they are relatively close---yet do not attain---the Singleton bound.
In~\cite{YY17},\cite{YY18}, a construction of optimal binary codes for two node failures was also presented based
upon ideas from EVENODD codes~\cite{DBLP:journals/tc/BlaumBBM95}.

Another approach for handling symmetric crisscross erasures
(in symmetric arrays) is by using symmetric rank-metric codes.
In~\cite{journals/jct/Schmidt10}, Schmidt presented
a construction of linear $[n\times n,k,d]$ symmetric binary
array codes with minimum rank~$d$, where $k=n(n{-}d{+}2)/2$
if $n{-}d$ is even, and $k=(n{+}1)(n{-}d{+}1)/2$ otherwise.
Such codes can correct any $d-1$ column or row erasures.
Hence, it is possible to use these codes to derive
$\rho$-node-failure-correcting codes while setting $d=2\rho{+}1$,
as the $\rho$ node failures translate into the erasure of
$\rho$ columns and $\rho$ rows. However, the redundancy of
these codes is $\binom{\rho}{2}$ symbols away from
the Singleton bound for symmetric crisscross erasures
(e.g., for $\rho = 2$, their redundancy is $2n$ while
the Singleton lower bound is $2n-1$).

In this paper we carry an algebraic approach such as the one presented in~\cite{DBLP:journals/tit/BlaumR93} in order to propose new constructions of binary codes over graphs. In Section~\ref{sec:defs}, we formally define codes over graphs and review several basic properties from~\cite{YY17,YY18b} that will be used in the paper. In Section~\ref{sec:two-node}, we present our optimal binary construction for two-node failures and its decoding procedure in Section~\ref{sec:decoding}. This construction is simpler than our optimal construction from~\cite{YY17},\cite{YY18b}. Then, in Section~\ref{sec:three-node}, we extend this construction for the three-node failures case. This new construction is only at most a single bit away from the Singleton bound, thereby outperforming the construction obtained from~\cite{journals/jct/Schmidt10}. Lastly, Section~\ref{sec:conc} concludes the paper.

\section{Definitions and Preliminaries}\label{sec:defs}		
For a positive integer $n$, the set $\{0,1,\ldots,n-1\}$ will be denoted by $[n]$ and for a prime power $q$, ${\Fq}$ is the finite field of size $q$. A linear code of length $n$ and dimension $k$ over $\Fq$ will be denoted by $[n,k]_q$ or $[n,k,d]_q$, where $d$ denotes its minimum distance. In the rest of this section, we follow the definitions of our previous work~\cite{YY17} for codes over graphs.
		
		A graph will be denoted by $G=(V_n,E)$, where $V_n=\{v_0,v_1,\ldots,v_{n-1}\}$ is its set of $n$ nodes (vertices) and $E\subseteq V_n\times V_n$ is its edge set. In this paper, we only study complete undirected graphs with self-loops, and in this case, the edge set of an undirected graph $G$ over an alphabet $\Sigma$ is defined by $E=\{(v_i,v_j)~|~(v_i,v_j)\in V_n\times V_n, i\geq j \}$, with a labeling function $L:V_n\times V_n\rightarrow \Sigma$. By a slight abuse of notation, every undirected edge in the graph will be denoted by $\langle v_i,v_j\rangle$ where the order in this pair does not matter, that is, the notation $\langle v_i,v_j \rangle$ is identical to the notation $\langle v_j,v_i \rangle$, and thus there are $\binom{n+1}{2}$ edges. We will use the notation $G=(V_n,L)$ for such graphs. For the rest of the paper, whenever we refer to a graph we refer to an undirected graph. 
		
		The \textit{\textbf{labeling matrix}} of an undirected graph $G=(V_n,L)$ is an $n\times n$ symmetric matrix over $\Sigma$ denoted by $A_G=[a_{i,j}]^{n-1,n-1}_{i=0,j=0}$, where $a_{i,j}= L\langle v_i,v_j \rangle$. We also use the \textit{\textbf{lower-triangle-labeling matrix}} of $G$ to be the $n\times n$ matrix $A_G'=[a'_{i,j}]^{n-1,n-1}_{i=0,j=0}$ such that $a'_{i,j}=a_{i,j}$ if $i\geq j$ and otherwise $a'_{i,j}=0$. 
%The \textit{\textbf{upper-triangle-adjacency matrix}} is defined similarly. 
	The \textit{\textbf{zero graph}} will be denoted by $G_{\textbf{0}}$ where for all $i,j\in[n]$, $a_{i,j}=0$. 
		
		\begin{comment}
		For $i\in[n]$, the \textit{neighborhood} of the $i$th node, denoted by $N_i$, is the set of edges connected to this node. Since we assumed the graph is complete, the neighborhood is simply the set $N_i=\{(v_i,v_j) | j\in[n] \}$.
		\end{comment}

		\begin{comment}
		The next example demonstrates the above definitions for undirected graphs. 
		\begin{example}
			Let $G$ be a complete undirected graph with self-loops over $\F_2$ and let $V_4=\{v_0,v_1,v_2,v_3\}$ be its node set. The graph $G$, its adjacency matrix $A_G$, and lower-triangle-adjacency matrix $A'_G$ are shown in Fig.~\ref{fig:graph example1}, where the edges $\langle  v_0,v_1 \rangle,\langle v_0,v_2\rangle,\langle v_0,v_3\rangle$ and $\langle v_1,v_2\rangle$ are labeled with $1$ and the rest of the edges are labeled with $0$.
			\begin{figure}[h!]
				\centering
				\includegraphics[width=80mm]{./figures/Example1.pdf}
				\caption{The undirected graph $G$, its adjacency matrix $A_G$, and its lower-triangle-adjacency matrix $A_G'$.}
				\label{fig:graph example1}
			\end{figure}	
		\end{example}
		
	\end{comment}
		
		Let $\Sigma$ be a ring and $G_1$ and $G_2$ be two graphs over $\Sigma$ with the same node set $V$. The operator $``+"$ between $G_1$ and $G_2$ over $\Sigma$, is defined by $G_1 + G_2 = G_3$, where $G_3$ is the unique graph satisfying $A_{G_1} + A_{G_2} = A_{G_3}$.
		Similarly, the operator~$``\cdot"$~between $G_1$ and an element $\alpha \in \Sigma$, is denoted by $\alpha\cdot G_1 = G_3$, where $G_3$ is the unique graph satisfying $\alpha \cdot A_{G_1} = A_{G_3}$.
		
%		\begin{definition}
	%		Let $V_n$ be the set of nodes $V_n =\{v_0,\ldots,v_{n-1}\}$. %A \textit{\textbf{code over graphs}} over $\Sigma$ of length $n$ and size $M$ is a set of undirected graphs $\mathcal{C}_{\cD}=\{G_{i} = (V_{n},L_{\cD_i})  |  i\in[M]\}$ over $\Sigma$, denoted by $\dc(n,M)_{\Sigma}$. Similarly 
		A \textit{\textbf{code over graphs}} over $\Sigma$ of length $n$ and size $M$ is a set of graphs $\mathcal{C}=\{G_{i} = (V_{n},L_{i})  |  i\in[M]\}$ over $\Sigma$, and it will be  denoted by $(n,M)_{\Sigma}$. In case that $\Sigma =\{0,1\}$, we simply use the notation $(n,M)$. The \textit{\textbf{dimension}} of a code over  graphs $\mathcal{C}$ is $k= \log_{|\Sigma|}M$ and the \textit{\textbf{redundancy}} is $r = {n+1 \choose 2} - k$. A code over  graphs $\mathcal{C}$ over a ring $\Sigma$ will be called \textit{\textbf{linear}} and will be denoted by $\uc[n,k]_\Sigma$ if for every $G_1,G_2 \in \mathcal{C}$ and $\alpha,\beta \in \Sigma$, $\alpha G_1 + \beta G_2 \in \mathcal{C}$.
			
			%A linear code over graphs will be called \textit{\textbf{systematic}} and will be denoted by $\cS\uc[n,k]_{\Sigma}$ if the first $k$ nodes contain the $\binom{k+1}{2}$ unmodified information symbols on their edges. The remaining ${n+1 \choose 2} -  {k+1 \choose 2}$ edges are called \textit{\textbf{redundancy edges}}. In this case we say that there are $k$ \textit{\textbf{information nodes}} and $r=n-k$ \textit{\textbf{redundancy nodes}}. The number of \textit{\textbf{information edges}} is $k_\cU={k+1 \choose 2}$ and the redundancy is $r_\cU = {n+1 \choose 2} -  {k+1 \choose 2}$. 
			
			% , and the rate is $R_\cU = {{k+1 \choose 2}}/{{n+1 \choose 2}}$. 
%		\end{definition}
		
%		\begin{definition}
%			Let $G=(V_n,L_\cU)$ be a graph. 
			
			The \textit{\textbf{neighborhood edge set}} of the $i$th node of an undirected graph $G=(V_n,L)$ is defined by $N_i = \{\langle v_i,v_j\rangle \ | \ j\in[n]  \}$, and it corresponds to the $i$th column and the $i$th row in the labeling matrix $A_{G}$.	
%			For $i\in[n]$, the \textit{\textbf{out-neighborhood edge set}}, \textit{\textbf{in-neighborhood edge set}}, of the $i$-th node is defined to be the set $$N_i^{\mathrm{out}} = \{(v_i,v_j)\ | \ j\in[n] \}, \  N_i^{\mathrm{in}} = \{(v_j,v_i)\ | \ j\in[n] \},$$ respectively, and the \textit{\textbf{neighborhood edge set}} of the $i$-th node is the set  $N_i = N_i^{\mathrm{out}}\cup N_i^{\mathrm{in}}$. Note that the $i$-th out-neighborhood, in-neighborhood edge set, corresponds to the $i$-th row, column, in an adjacency matrix $A_{G}$, respectively, and the $i$-th neighborhood edge set is the union of the $i$-th column and the $i$-th row in the adjacency matrix. Similarly, the neighborhood edge set of the $i$-th node of an undirected graph $G=(V_n,L_\cU)$ is defined by $N_i = \{\langle v_i,v_j\rangle \ | \ j\in[n]  \}$, which corresponds to the $i$-th column and row in an adjacency matrix $A_{G}$.			
			The \textit{\textbf{node failure}} of the $i$th node is the event in which all the edges in the neighborhood set of the $i$th node, i.e. $N_i$, are erased. We will also denote this edge set by $F_i$ and refer to it by the \textit{\textbf{failure set}} of the $i$th node. A code over graphs is called a $\rho$\emph{\textbf{-node-erasure-correcting code}} if it can correct any failure of at most $\rho$ nodes in each of its graphs.
%		\end{definition}
		
		\begin{comment}
		\begin{example}\label{ex:SECGC}
		The following graph code, given in Fig.~\ref{fig:graph code example2}, is a single-erasure-correcting graph code. Note that the graph code in Example~\ref{ex:linear_GC} is not a single-erasure-correcting graph code, since the erasure of the first node $v_0$ in graphs 1 and 2 will result with the same graph, and hence we won't be able to reconstruct the erased node.
		%	A linear binary systematic single-erasure-correcting-graph code of length $3$, with rate of $R_U = \frac{1}{2}$, redundancy $r_U = 3$ and $r=1$ redundancy node. The blue nodes are the information nodes and a purple node is dundancy node.
		\begin{figure}[h!]
		\centering
		\includegraphics[width=70mm]{./figures/Graph_Example2}
		\caption{A $\cS\gc[3,2]$ single-erasure-correcting-graph code.}
		\label{fig:graph code example2}
		\end{figure}
		\end{example}
		\end{comment}
		
		As discussed in~\cite{YY17,YY18,YY18b}, according to the Singleton bound, the minimum redundancy $r$ of any $\rho$-node-erasure-correcting code of length $n$, satisfies
		\begin{align}
			& r\geq \binom{n+1}{2}-\binom{n-\rho+1}{2} =  n\rho-\binom{\rho}{2},\label{eq:singleton_bound1}
		\end{align}
		 and a code over graphs which satisfies this inequality with equality is called \textit{\textbf{optimal}}. %Every optimal code according to the bound (\ref{eq:singleton_bound1}) has also a systematic form and thus for systematic codes over graphs the number of redundancy nodes is at least $\rho$. 
		 It was also observed in~\cite{YY17,YY18,YY18b} that for all $n$ and $\rho$, an optimal $\rho$-node-erasure-correcting code exists over a field of size at least $\Theta(n^2)$, and thus the goal is to construct such codes over smaller fields, and ideally over the binary field. 
		 
%		  from an $[\binom{n+1}{2}, \binom{n-\rho+1}{2},n\rho  -\binom{\rho}{2}+1]$ MDS code. However, in both cases the field size of the code over graphs will be at least $\Theta(n^2)$. Our goal in this work is to construct $\rho$-node-erasure-correcting codes over smaller fields. When possible, we seek the field size to be binary and in any event at most $\cO(n)$. 

We conclude this section with reviewing the definition of a distance metric over graphs from~\cite{YY18b} and its connection to the construction of codes correcting node failures. Let $G=(V_n,L)$ be a graph and let $E$ be a set of all nonzero labeled edges of $G$, i.e., $E = \{e\in V_n\times V_n ~|~ L(e)\neq 0 \}$. A \textit{\textbf{vertex cover}} $W$ of $G$ is a subset of $V_n$ such that for each $\langle v_i,v_j\rangle \in E$ either $v_i\in W$ or $v_j\in W$. The \textit{\textbf{graph weight}} of $G$ is defined by $$w(G) = \min_{W \textmd{ is a vertex cover of $G$}}\{|W|\},$$
and the \textit{\textbf{graph distance}} between two graphs $G_1,G_2$ will be denoted by $d(G_1,G_2)$ where it holds that $d(G_1,G_2)=w(G_1-G_2)$. It was proved in~\cite{YY18b} that this graph distance is a metric. The \textit{\textbf{minimum distance}} of a code over graphs $\cC$, denoted by $d(\cC)$, is the minimum graph distance between any two distinct graphs in $\cC$, that is
	$$d(\cC) = \min_{G_1\neq G_2~ G_1,G_2 \in \cC \textmd{}}\{d(G_1,G_2)\},$$
	and in case the code is linear
	$d(\cC) = \min_{G\in \cC \textmd{}, G\neq G_{\textbf{0}}}\{w(G)\} $.
	Lastly, we state the following theorem from~\cite{YY18b} that establishes the connection between the graph distance and the node-erasure-correction capability. 
\begin{theorem}\label{th:mid_dist}
 A linear code over graphs $\cC$ is a $\rho$-node-erasure-correcting code if and only if its minimum distance satisfies $d(\cC) \geq \rho+1$.
\end{theorem}

Let $n\geq 2$ be a prime number. Denote by $\cR_n$ the ring of polynomials of degree at most $n-1$ over $\F_2$. It is well known that $\cR_n$ is isomorphic to the ring of all polynomials in $\F_2[x]$ modulo $x^n-1$. Denote by $M_n(x)\in \cR_n$ the polynomial $M_n(x) = \sum^{n-1}_{\ell=0}x^\ell$ over $\F_2$, where it holds that $M_n(x)(x+1) = x^n-1$ as a multiplication of polynomials over $\F_2[x]$. To avoid confusion in the sequel, since we are using only polynomials over $\F_2$, the notation $x^\ell+1$ for all $ \ell \in [n]$, will refer to a polynomial in $\cR_n$ and for $\ell = n$, we will use the notation $x^n-1$. It is well known that for all $ \ell \in [n]$ it holds that
$$\gcd(x^\ell+1,x^n-1) = x^{\gcd(\ell,n)}+1 = x+1,$$
and since $M_n(x)(x+1) = x^n-1$ it can be verified that
\begin{align}\label{Mn_prop1}
\gcd(x^\ell+1,M_n(x)) =1.
\end{align}
Notice also that when $2$ is primitive in $\Z_n$, the polynomial $M_n(x)$ is irreducible~\cite{P-MDS}. The last important and well known property we will use for polynomials over $\F_2$ is that for all $k=2^j$, $j\in \N$ it holds that $1+x^{sk}=(1+x^s)^k$. The notation $\langle a \rangle_n$ will be used to denote the value of $(a \bmod n)$.

\begin{comment}
We use the following result of Gustavson~\cite{GostavsonMatrix} for our efficient decoding procedure of the two erasure case: 
\begin{theorem}\label{fastmatrix}
	Given a field $\mathbb{F}$ and two Matrices $A\in \mathbb{F}^{k\times t}, B\in \mathbb{F}^{t\times s}$ where $t,k,s\leq n$, the matrix $AB$ cab be computed in $O(mn)$ time where $m$ is a bound on the number of non zero entries of both $A$ and $B$.
\end{theorem}
\end{comment}

	\begin{figure*}[t!]	
		\centering
		\subfigure[Neighborhood-Parity Constraints]{\includegraphics[width=80mm]{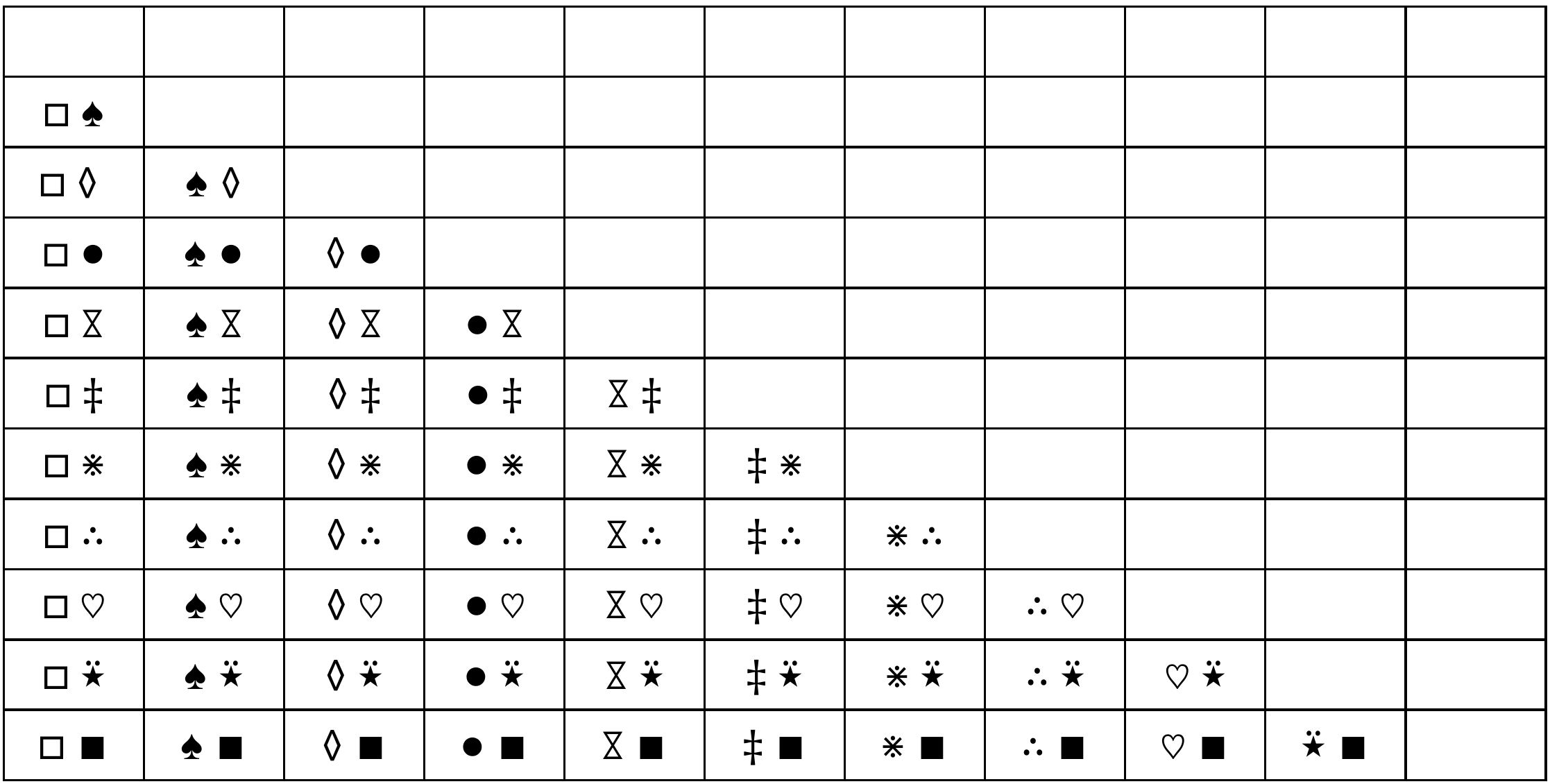}}~~~~~~~~~~~~
		\subfigure[Slope-One-Diagonal-Parity Constraints]{\includegraphics[width=80mm]{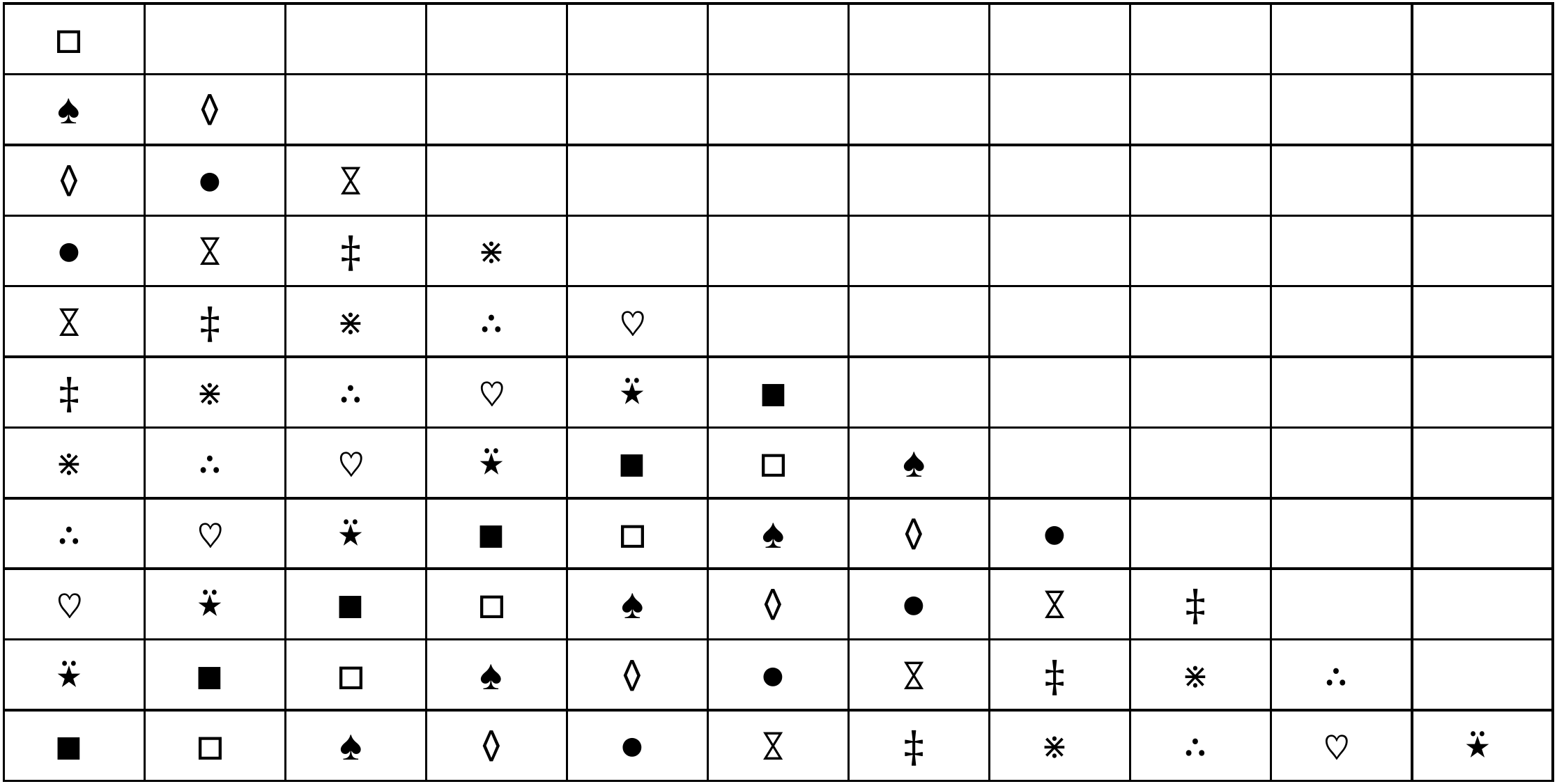}}
		\caption{The constraints over undirected graphs, represented on the lower-triangle-labeling matrix.} \label{fig:ex2}	
	\end{figure*}
	
\section{Optimal Binary Double-Node-Erasure-Correcting Codes}\label{sec:two-node}
In this section we present a family of optimal binary linear double-node-erasure-correcting codes with $n$ nodes, where $n$ is a prime number. 

Remember that for $i\in[n]$ the $i$th neighborhood set of the $i$th node is $N_i = \{\langle v_i,v_j\rangle \ | \ j\in[n]  \}$. %For notational purposes, in this section we denote the edge $\langle v_i,v_j\rangle$ by $e_{i,j}$, and here . 
Let $n\geq 2$ be a prime number and let $G = (V_n,L)$ be a graph with $n$ vertices. For $h\in [n]$ we define the neighborhood of the $h$th node without itself self-loop by
\begin{equation}\label{S_h}
S_{h} = \big\{\langle v_h,v_\ell \rangle~|~\ell\in [n], h\neq \ell\big\}.
\end{equation}
We also define for $m\in[n]$, the $m$th diagonal set by
\begin{equation}\label{D_m}
D_{m} = \{\langle v_k,v_{\ell} \rangle|k,\ell \in[n],  \langle k + \ell \rangle_n = m \}.
\end{equation}
The sets $S_{h}$ for $h\in[n]$ will be used to represent parity constraints on the neighborhood of each node and similarly the sets $D_{m}$ for $m\in[n]$ will be used to represent parity constraints on the diagonals with slope one in the labeling matrix $A_G$. We state that for all $m\in [n]$, the size of $D_m$ is $\frac{n+1}{2}$. 
This holds since in each neighborhood $N(v_i)$, there is only  a single edge which belongs to $D_m$, which is the edge $\langle v_i,v_{\langle m-i \rangle_n}\rangle$. 
Another important observation is that $D_m$ contains only a single self-loop which is the edge $\langle v_{\ipn{m\cdot 2^{-1}}},v_{\langle m\cdot 2^{-1} \rangle_n} \rangle$. %Hence, each diagonal set $D_m$ is a perfect matching of the nodes, and in the graphs we discuss the size of a perfect matching is exactly $\frac{n+1}{2}$.

\begin{example}
	In Fig.~\ref{fig:ex2} we demonstrate the sets $S_h$ and $D_m$, where $h,m\in[11]$, of a graph $G = (V_{11}, L)$ on its lower-triangle-labeling matrix $A_G'$.
\end{example}

We introduce one more useful notation for graphs. Let $G = (V_n,L)$ be a graph. For $i \in [n]$ we denote the \textit{neighborhood-polynomials} of $G$ to be 
\begin{align*}
a'_i(x) = e_{i,0}+e_{i,1}x+e_{i,2}x^2+\dots + e_{i,n-1}x^{n-1},
\end{align*}
where for $i,j\in[n]$, $e_{i,j} = a_{i,j} = L \langle v_i,v_j \rangle $. We also denote the \textit{neighborhood-polynomial without self-loops} of $G$ to be
\begin{align*}
& a_i(x) =a'_i(x) - e_{i,i}x^i.
\end{align*}

We are now ready to present the construction of optimal double-node-erasure-correcting codes.
\begin{Construction}\label{const:double}
Let $n\geq 2$ be a prime number. The code over graphs $\cC_2$ is defined as follows,
	\begin{equation*}
	\cC_2 = \left\{G = (V_n,L)  \middle|
	\begin{array}{cc}
	(a)& \sum_{\langle v_i,v_j \rangle\in S_h} e_{i,j} = 0, h \in  [n]  \\ 
	(b)&  \sum_{\langle v_i,v_j \rangle\in D_m} e_{i,j}=0, m\in[n]
	\end{array}
	\right\}.
	\end{equation*}
\end{Construction}
Note that for \emph{any} graph $G$ over the binary field, it holds that
%	Since $\cC_2$ is a binary code, it holds that
	\begin{align}\label{double_red}
	\sum_{h\in[n]}\sum_{\langle v_i,v_j \rangle\in S_h} e_{i,j}  = \sum_{h=0}^{n-1}\sum_{\ell=0,\ell\neq h}^{n-1}e_{h,\ell}  
	=  2 \sum_{h=0}^{n-1}\sum_{\ell=0}^{h-1}e_{h,\ell}  = 0.
	\end{align}
	Therefore the code $\cC_2$ has at most $2n-1$ linearly independent constraints which implies that its redundancy is at most $2n-1$.  Since we will prove in Theorem~\ref{th:double} that $\cC_2$ is a double-node-correcting codes, according to the Singleton bound we get that the redundancy of the code $\cC_2$ is exactly $2n-1$, and thus it is an optimal code.
	
	According to Theorem~\ref{th:mid_dist}, in order to prove that $\cC _2$ is a double-node-erasure-correcting code, we need to show that $d(\cC_2) \geq 3$, that is, for every $G\in \cC _2$, $w(G)\geq 3$. This will be proved in the next theorem.

\begin{theorem}\label{th:double}
For all prime number $n$, the code $\cC_2$ is an optimal double-node-erasure-correcting code.
\end{theorem}
\begin{IEEEproof}
Assume in the contrary that $d(\cC_2) \leq 2$ and let $G\in\cC_2, G\neq G_{\textbf{0}}$ be a nonzero graph such that $w(G) =2$ (a similar proof will hold in case $w(G) =1$). Since  $w(G) =2$, the graph $G$ has a vertex cover of size 2, that is, all its nonzero edges are confined to the neighborhoods $N_i,N_j$ of some two nodes $v_i,v_j$. By symmetry of the graph, it suffices to prove the above property for the case where the two nodes are $v_0,v_i$ for some $i\neq 0$. During the proof, we assume that $a_i(x)$, for $i\in[n]$ are the neighborhood polynomials of the graph $G$.
We first prove the following two claims. 
\begin{claim}\label{cl1}
The following properties hold on the graph $G$:
\begin{enumerate}
    \item\label{cl1a} For all $h\in [n]\setminus \{0,i\}$, $e_{h,0} +e_{h,i}=0$.
    \item\label{cl1b} For all $h\in [n]\setminus \{i\}$, $e_{0,h}+e_{i,\langle h-i\rangle_n}=0$.
    \item\label{cl1c} $e_{0,i}=0$.
\end{enumerate}
\end{claim}
\begin{IEEEproof} 
\begin{enumerate}
\item According to the neighborhood constraint $S_h$  for all $h\in [n]\setminus \{0,i\}$, we have that $$ 0 =\sum\limits _{\langle v_h,v_{\ell}\rangle \in S_h} e_{h,\ell} = \sum_{\ell=0,\ell\neq h}^{n-1}e_{h,\ell} = e_{h,0} +e_{h,i},$$ and since $e_{h,\ell} =0$ for all $\ell\in [n]\setminus \{0,i\}$, we get that $e_{h,0} +e_{h,i}=0$.

\item For $h\in [n]\setminus \{i\}$, denote the set $D_h\backslash \set{ \langle v_0,v_h \rangle,\langle v_i,v_{\langle h-i \rangle_n }\rangle }$ by $D'_h$. Therefore, we have that 
\begin{align*}
&0 =\sum\limits _{\langle  v_\ell,v_{\langle h-\ell \rangle_n} \rangle \in D_h} e_{\ell,\langle h-\ell \rangle_n} \\
&=\sum\limits _{\langle  v_\ell,v_{\langle h-\ell \rangle_n} \rangle \in D'_h} e_{\ell,\langle h-\ell \rangle_n} + e_{0,h}+e_{i,\langle h-i \rangle_n},
\end{align*}
and since $e_{s,\ell} =0$ for all $\langle v_s,v_{\ell} \rangle \in  D'_h$, we get that $e_{0,h} +e_{i,\langle h-i\rangle_n}=0$.

\item According to the diagonal constraint $D_i$ we get that 
\begin{align*}
& 0=\sum\limits _{ \langle v_s,v_{\ell} \rangle \in D_i} e_{s,\ell}  = e_{0,i} + \sum\limits _{\langle  v_s,v_{\ell} \rangle \in D_i\setminus \set{\langle v_0,v_i \rangle }} e_{s,\ell},
\end{align*}
and since $e_{s,\ell} =0$ for all
$\langle v_s,v_{\ell} \rangle \in D_i\setminus \set{\langle v_0,v_i \rangle }$, we get that $e_{0,i}=0$.
\end{enumerate}

\end{IEEEproof}

\begin{claim}\label{cl2}
The following properties hold on the graph $G$: %\in \cC _2$:
\begin{enumerate}
    \item\label{cl2a} For all $h\in[n]$, $a_h (1)=0$.
    \item\label{cl2b} $a_0(x)+a_i(x)=0$.
%    \item\label{cl2c} $a_0(x)+a_i(x)x^i=e_{0,0}+e_{i,i}x^{\langle 2i\rangle_n}$.
        \item\label{cl2c} $a_0(x)+a_i(x)x^i \equiv e_{0,0}+e_{i,i}x^{2i}   (\bmod x^n-1)$.
\end{enumerate}
\end{claim}
\begin{IEEEproof}
\begin{enumerate}
\item By the definition of the neighborhood constraints, for all $h\in[n]$, $S_h = \{\langle v_h,v_\ell \rangle~ |~\ell\in [n], h\neq \ell \} $, and therefore
\begin{gather*}
a_h(1)=\sum_{\ell=0,\ell\neq h} ^{n-1} e_{h,\ell}=\sum_{\langle  v_h,v_{\ell} \rangle\in S_h} e_{h,\ell}=0.
\end{gather*}
	\item % By definition, 
	\begin{align*}
	& a_0(x) + a_i(x) =  \\
	& =  e_{0,0}+e_{i,i}x^i  + \sum^{n-1}_{\ell = 0}e_{0,\ell}x^{\ell}+\sum^{n-1}_{\ell = 0}e_{i,\ell}x^{\ell} \\
	& = e_{0,0}+e_{i,i}x^i  + \sum^{n-1}_{\ell = 0}\Big(e_{0,\ell}+e_{i,\ell} \Big)x^{\ell}, \\
	& \overset{(a)}{=} e_{0,0}+e_{i,i}x^i    + \Big(e_{0,0}+e_{i,0} \Big) + \Big(e_{0,i}+e_{i,i} \Big)x^i \\
	&= e_{i,0}(1+x^i) \overset{(b)}{=} 0 ,
	\end{align*}
	where Step (a) holds since by Claim~\ref{cl1}(\ref{cl1a}) for all $\ell\in[n]\setminus \{0,i \}$, $e_{0,\ell}+e_{i,\ell} = 0$ and Step (b) holds
	since by Claim~\ref{cl1}(\ref{cl1c}), $e_{i,0}=0$. 

	\item %By definition,
	\begin{align*}
	& a_0(x) + a_i(x)x^i = \\
	&= e_{0,0}+e_{i,i}x^{2i}   + \sum^{n-1}_{\ell = 0}e_{0,\ell}x^{\ell}+\sum^{n-1}_{\ell = 0}e_{i,\ell}x^{\ell+i} \\
	&\equiv e_{0,0}+e_{i,i}x^{2i}  + \sum^{n-1}_{\ell = 0}e_{0,\ell}x^{\ell}+\sum^{n-1}_{\ell = 0}e_{i,\langle \ell -i \rangle_n}x^{\ell} (\bmod x^n-1)\\
	&\equiv e_{0,0}+e_{i,i}x^{2i}   + \sum^{n-1}_{\ell = 0}\Big(e_{0,\ell}+e_{i,\langle \ell -i \rangle_n}\Big)(\bmod x^n-1)\\
	&  \overset{(a)}{\equiv} e_{0,0}+e_{i,i}x^{2i}  + \Big(e_{0,i}+e_{i,0}\Big)x^i  (\bmod x^n-1)\\
	& \equiv e_{0,0}+e_{i,i}x^{2i}   (\bmod x^n-1),
	\end{align*}
	where Step (a) holds since by Claim~\ref{cl1}(\ref{cl1b}) for all $\ell\in[n]\setminus \{i  \}$, $e_{0,\ell}+e_{i,\langle \ell -i \rangle_n}=0$.
\end{enumerate}
 \end{IEEEproof}

The summation of the equations from Claims~\ref{cl2}(\ref{cl2b}) and~\ref{cl2}(\ref{cl2c})  results with
$$
a_i(x)(1+x^i) \equiv e_{0,0}+e_{i,i}x^{2i} (\bmod x^n-1).
$$
It holds  that $e_{0,0}=e_{i,i}$ by applying $x=1$ in the last equation. Assume that $e_{0,0}=e_{i,i}=1$, so we get that 
$$a_i(x)(1+x^i)\equiv 1+x^{2i}(\bmod x^n-1).$$
Since $1+x^{2i} = (1+x^i)^2$, it holds that 
$$(1+x^i)( 1+x^i+ a_i(x)) \equiv 0 (\bmod x^n-1).$$
%we get by moving the left side to the right side and taking $(1+x^i)$ as a common multiplier that 
%\begather{(1+x^i)(a_i(x)+1+x^i)=0 \bmod x^n-1}
Denote by $p(x)$ the polynomial $p(x)= 1+x^i + a_i(x)$, and  since $p(1)=0$, it holds that $1+x|p(x)$.  
As stated in~\eqref{Mn_prop1}, it holds that
$\gcd(x^{i}+1,M_n(x)) =1,$
and since 
$$(1+x^i)p(x) = (x^n-1)s(x) = M_n(x)(x+1)s(x)$$
for some polynomial $s(x)$ over $\F_2$, we deduce that $M_n(x)|p(x)$. Therefore we get that $x^n-1|p(x)$, however $p(x)\in \cR_n$, and so we deduce that $p(x) = 0$, that is, $a_i(x) = 1+x^i$. This results with a contradiction since the coefficient of $x^i$ in $a_i(x)$ is 0. Thus $e_{0,0}=e_{i,i}=0$ and 
$$a_i(x)(1+x^i)\equiv 0(\bmod x^n -1).$$
%Thus we conclude that that $p(x)=0$, which in particular means that $a_i(x)=1+x^i$. However, this is a contradiction since the coefficient of $x^i$ in $a_i(x)$ is 0. Thus $e_{0,0}=e_{i,i}=0$ and 
%\begather{a_i(x)(1+x^i)=0\bmod x^n -1}
%Now assume that $a_i(x)\neq 0$,
Notice that $a_i(x)\in \cR_n$ and by Claim~\ref{cl2}(\ref{cl2a}) it also holds $a_i(1) = 0 $. Since $\gcd(x^{i}+1,M_n(x)) =1$, we derive that $x^n-1|a_i(x)$ and since $a_i(x)\in \cR_n$, we immediately get that $a_i(x) = 0$. Finally, from Claim~\ref{cl2}(\ref{cl2b}) we get also that $a_0(x)=0$ and together we get that $G= G_{\textbf{0}}$, which is a contradiction. This completes the proof.
\end{IEEEproof}

\section{Decoding of the Double-Node-Erasure-Correcting Codes}\label{sec:decoding}
In Section~\ref{sec:two-node}, we proved that the code $\cC_2$ can correct the failure of any two nodes in the graph. 
Note that whenever two nodes fail, the number of unknown variables is $2n-1$, and so a naive decoding solution for the code $\cC_2$ is to solve the linear equation system of $2n-1$ constraints with the $2n-1$ variables. However, the complexity of such a solution will be $O(n^\omega)$, where it is only known that $2\leq \omega \leq 2.37286$ as it requires the inversion of a $(2n-1)\times (2n-1)$ matrix~\cite{LG14}. Our main goal in this section is a decoding algorithm for $\cC_2$ of time complexity $\Theta(n^2)$. Clearly, this time complexity is optimal since the complexity of the input size of the graph is $\Theta(n^2)$.

Throughout this section we assume that $G$ is a graph in the code $\cC_2$ and $a_\ell(x)$ for $\ell\in[n]$ are its neighborhood polynomials. We also assume that the failed nodes are $v_0,v_i$. First, we define the following two polynomials $S_1(x),S_2(x)\in\cR_n$, which will be called the \emph{syndrome polynomials}
%Given that the failed vertices are $v_0,v_i$ we define the following  two polynomials:
\begin{gather*}
S_1(x)=\sum\limits_{\ell=1,\ell\neq i} ^{n-1} a_\ell(x), \\
S_2(x) \equiv \sum\limits_{\ell=1,\ell\neq i} ^{n-1} a_\ell(x)x^\ell (\bmod x^n-1).
\end{gather*}
Note that if no nodes have failed in the graph $G$, then we can easily compute both of these polynomials since we know the values of all the edges. However in case that $v_0,v_i$ both failed this becomes a far less trivial problem. However, using several properties, that will be proved in this section, we will prove that %for any graph in the code (i.e., codeword) 
it is still possible to compute $S_1(x)$ entirely, and compute all the coefficients of $S_2(x)$ but the ones of $x^0$ and $x^{\ipn{2i}}$, even though the nodes $v_0,v_i$ failed.

Our goal in this section is to prove the following theorem.
\begin{theorem}\label{Th4}
There exists an efficient decoding procedure to the code $\cC _2$ given any two node failures. Its complexity is $\Theta(n^2)$, where $n$ is the number of nodes.
\end{theorem}

Before we present the proof of Theorem~\ref{Th4}, we prove a few properties of the code that will help up to present the decoding procedure. %For the rest of this section $G$ is a graph in the code $\cC_2$ and $a_i(x)$, for $i\in[n]$ are its neighborhood polynomials. 
\begin{claim}\label{cl3}
It holds that 
\begin{enumerate}
\item\label{cl3a} $\sum\limits _{\ell=0} ^{n-1} a_\ell(x)=0$.
\item\label{cl3b} $\sum\limits _{\ell=0} ^{n-1} a_\ell(x)x^\ell = 0$.
\end{enumerate}
\end{claim}
\begin{IEEEproof}
\begin{enumerate}
\item  The coefficient of some $x^\ell$ is the sum of all edges $e_{k,\ell}$, where $k\neq \ell$, and so we get that 
$$\sum\limits _{\ell=0} ^{n-1} a_\ell(x)=\sum\limits _{\ell=0} ^{n-1}(\sum\limits _{k=0,k\neq \ell} ^{n-1} e_{k,\ell}x^k) =\sum\limits _{k=0} ^{n-1} ( \sum\limits _{\ell=0,\ell\neq k} ^{n-1} e_{k,\ell}) x^k=0,$$
where the second transition is a result of changing the order of the sum and the last equality holds by the neighborhood constraint on the $k$th node. % applying Claim \ref{cl2}(\ref{cl2a}) on $S_k$ for all $k\in [n]$.

\item Note that 
\begin{align*}
\sum\limits _{\ell=0} ^{n-1} a_\ell(x)x^\ell & = \sum\limits _{\ell=0} ^{n-1} (\sum\limits _{k=0,k\neq \ell} ^{n-1} e_{k,\ell}x^k)x^\ell \\
& = \sum\limits _{\ell=0} ^{n-1} \sum\limits _{k=0,k\neq \ell} ^{n-1} e_{k,\ell}x^{\ell+k}\\ 
&\overset{(a)}{=}
\sum\limits _{\ell=0} ^{n-1} \sum\limits _{k=0} ^{\ell-1} e_{k,\ell}x^{\ell+k}+\sum\limits _{\ell=0} ^{n-1} \sum\limits _{k=\ell+1} ^{n-1} e_{k,\ell}x^{\ell+k}\\
&\overset{(b)}{=}\sum\limits _{\ell=1} ^{n-1} \sum\limits _{k=0} ^{\ell-1} e_{k,\ell}x^{\ell+k}+\sum\limits _{k=1} ^{n-1} \sum\limits _{\ell=0} ^{k-1} e_{k,\ell}x^{\ell+k} \\
&\overset{(c)}{=}\sum\limits _{\ell=1} ^{n-1} \sum\limits _{k=0} ^{\ell-1} e_{k,\ell}x^{\ell+k}+\sum\limits _{\ell=1} ^{n-1} \sum\limits _{k=0} ^{\ell-1} e_{k,\ell}x^{\ell+k}=0.
\end{align*}
Step (a) holds by splitting the sum, Step (b) is a result of changing the summation order in the second sum and noticing that in the first sum the $\ell=0$ iteration is empty, and lastly in Step (c) we simply changed the variables $k,\ell$ with each other in the second sum.
\end{enumerate}
\end{IEEEproof}
\begin{comment}
\begin{claim}\label{cl4}
It holds that $\sum\limits _{\ell=0} ^{n-1} a_\ell(x)x^\ell = 0$.
\end{claim}
\begin{IEEEproof} 
Note that 
\begin{align*}
\sum\limits _{\ell=0} ^{n-1} a_\ell(x)x^\ell & = \sum\limits _{\ell=0} ^{n-1} (\sum\limits _{k=0,k\neq \ell} ^{n-1} e_{k,\ell}x^k)x^\ell \\
& = \sum\limits _{\ell=0} ^{n-1} \sum\limits _{k=0,k\neq \ell} ^{n-1} e_{k,\ell}x^{\ell+k}\\ 
&\overset{(a)}{=}
\sum\limits _{\ell=0} ^{n-1} \sum\limits _{k=0} ^{\ell-1} e_{k,\ell}x^{\ell+k}+\sum\limits _{\ell=0} ^{n-1} \sum\limits _{k=\ell+1} ^{n-1} e_{k,\ell}x^{\ell+k}\\
&\overset{(b)}{=}\sum\limits _{\ell=1} ^{n-1} \sum\limits _{k=0} ^{\ell-1} e_{k,\ell}x^{\ell+k}+\sum\limits _{k=1} ^{n-1} \sum\limits _{\ell=0} ^{k-1} e_{k,\ell}x^{\ell+k} \\
&\overset{(c)}{=}\sum\limits _{\ell=1} ^{n-1} \sum\limits _{k=0} ^{\ell-1} e_{k,\ell}x^{\ell+k}+\sum\limits _{\ell=1} ^{n-1} \sum\limits _{k=0} ^{\ell-1} e_{k,\ell}x^{\ell+k}=0.
\end{align*}
Step (a) holds by separating the sum, Step (b) is a result of changing the summation order in the second sum and noticing that in the first sum the $\ell=0$ iteration is empty, and lastly in Step (c) we simply changed the variables $k,\ell$ with each other in the second sum.
\end{IEEEproof}
\end{comment}
As an immediate result of Claim~\ref{cl3}, we get the following corollary.
\begin{corollary}\label{cor4}
It holds that \begin{align*}
& S_1(x)=a_0(x)+a_i(x), \\
& S_2(x)\equiv a_0(x)+a_i(x)x^i (\bmod{x^n-1}).
\end{align*}
\end{corollary}
\begin{IEEEproof}
According to Claim~\ref{cl3} we get that
\begin{align*}
	S_1(x)&=\sum\limits_{\ell=1,\ell\neq i} ^{n-1} a_\ell(x)  \\
	&= a_0(x) + a_i(x) + \sum\limits_{\ell=0} ^{n-1} a_\ell(x) \\
	& =  a_0(x) + a_i(x),
\end{align*}
and also,
\begin{align*}
S_2(x)&\equiv\sum\limits_{\ell=1,\ell\neq i} ^{n-1} a_\ell(x)x^{\ell} (\bmod{x^n-1}) \\
&\equiv a_0(x) + a_i(x)x^i + \sum\limits_{\ell=0} ^{n-1} a_\ell(x)x^{\ell}  (\bmod{x^n-1})\\
& \equiv  a_0(x) + a_i(x)x^i  (\bmod{x^n-1}).
\end{align*}
\end{IEEEproof}

Now we show that it is possible to compute $S_1(x)$, and almost compute $S_2(x)$ as explained above. 
\begin{claim}
         Given the two node failures $v_0,v_i$, it is possible to exactly compute the polynomial $S_1(x)$.
\end{claim}\label{cl5}
\begin{IEEEproof}
Let us consider the coefficient of $x^k$ in $S_1(x)$ for all $k\in[n]$. For each $a_\ell(x), \ell\in [n]\setminus \{k\}$ in the sum, the edge $e_{\ell,k}$ is added to the coefficient of $x^k$, and so we get that
\begin{align*}
S_1(x) &=\sum\limits _{\ell=1,\ell\neq i} ^{n-1} a_\ell(x)= \sum\limits _{\ell=1,\ell\neq i} ^{n-1}(\sum\limits_{k=0,k\neq \ell} ^{n-1} e_{k,\ell}x^k)\\
& =  \sum\limits _{k=0}  ^{n-1}(\sum\limits_{\ell=1,\ell\neq k,i} ^{n-1} e_{k,\ell})x^k,
\end{align*}
where in the last transition the summation order has been changed. For any $k\neq 0,i$, we can compute the coefficient of $x^k$ since we know all the edges in the sum. In case that $k=0$, we get that the coefficient of $x^0$ is 
$$\sum\limits_{\ell=1,\ell\neq i} ^{n-1} e_{\ell,0}=e_{i,0},$$
by the constraint $S_0$. For $k=i$, we get that the coefficient of $x^i$ is
$$\sum\limits_{\ell=1,\ell\neq i} ^{n-1} e_{\ell,i}=e_{0,i},$$
by the constraint $S_i$. Lastly, we know the value of $e_{0,i}$ by the diagonal constraint $D_i$, and therefore we know all the coefficients in $S_1(x)$.
\end{IEEEproof}

\begin{claim}
It is possible to compute all of the coefficients of the polynomial $S_2(x)$ except for the coefficients of $x^0$ and $x^{\ipn{2i}}$. Furthermore, the coefficients of these monomials are $e_{0,0}$ and $e_{i,i}$, respectively. 
\end{claim}
\begin{IEEEproof}
According to the definition of the polynomial $S_2(x)$ and Corollary~\ref{cor4} we know that $$S_2(x) \equiv a_0(x)+a_i(x)x^i (\bmod x^n-1),$$ which implies that
\begin{align*}
S_2(x) &\equiv e_{0,0} + e_{i,i}x^{\ipn{2i}} + \sum\limits _{\ell=0} ^{n-1} e_{0,\ell} x^\ell +\sum\limits _{\ell=0}^{n-1} e_{i,\ell} x^{\ell +i}  (\bmod x^n-1) \\
&\equiv e_{0,0} + e_{i,i}x^{\ipn{2i}} + \sum\limits _{\ell=0} ^{n-1} e_{0,\ell} x^\ell +\sum\limits _{\ell=0}^{n-1} e_{ i ,\langle \ell - i \rangle_n} x^{\ell} (\bmod x^n-1) \\
&\equiv e_{0,0} + e_{i,i}x^{\ipn{2i}} + \sum\limits _{\ell=0} ^{n-1}\Big( e_{0,\ell}  + e_{ i ,\langle \ell - i \rangle_n} \Big) x^{\ell}  (\bmod x^n-1).
\end{align*}
Notice that for all $\ell \in [n]$, all the values of the edges in the set $D'_{\ell} = D_{\ell}\setminus \{\langle v_0,v_\ell \rangle , \langle v_i,v_{\langle \ell -i \rangle_n \rangle } \}$ are known, and therefore, according to the diagonal constraint $D_\ell$ we get that the value of $e_{0,\ell}+e_{i,\ipn{\ell -i}}$ is calculated by
$$e_{0,\ell}+e_{i,\ipn{\ell -i}} = \sum_{\langle v_k,v_{\langle \ell - k \rangle_n \rangle} \in D'_{\ell}}e_{k,\langle \ell - k \rangle_n }.$$ 
 Finally, the only coefficients in this polynomial that we can not compute are the coefficients of $x^0$ and $x^{\ipn{2i}}$ which are $e_{0,0}$ and $e_{i,i}$, respectively. % as required and the proof is complete.

\end{IEEEproof}

\begin{claim}
         Given the values of $e_{0,0},e_{i,i}$, we can compute the polynomials $a_0(x)$ and $a_i(x)$, i.e., decode the failed nodes $v_0,v_i$.
\end{claim}
\begin{IEEEproof}
Assume that the values of $e_{0,0},e_{i,i}$ are known. This implies that we can compute exactly the polynomials $S_1(x)$ as well as $S_2(x)$ and let us denote
$$S_1(x)+S_2(x)\equiv\sum\limits _{k=0}^{n-1} s_{k}x^k (\bmod x^n -1),$$
that is, the coefficients $s_k$ for $k\in[n]$ are known. According to Corollary~\ref{cor4} we have that
\begin{align*}
& S_1(x)=a_0(x)+a_i(x), \\
 & S_2(x)\equiv a_0(x)+a_i(x)x^i(\bmod x^n -1).
\end{align*}
Adding up these two equations results with
$$S_1(x)+S_2(x)\equiv a_i(x)+a_i(x)x^i(\bmod x^n -1).$$
Thus, we get the following $n$ equations with the $n$ variables $e_{i,k}$ for $k\in[n]$. For all $k\in [n]\backslash \set{i,\ipn{2i}}$ we get the equation
$$e_{i,k}+e_{i,\ipn{k-i}}= s_k,$$
for $k=i$ we get the equation
$$e_{i,0}=s_i,$$
and lastly for $k=\ipn{2i}$ we get the equation
$$e_{i,\ipn{2i}}=s_{\ipn{2i}}.$$
We know that this linear system of equations has a single solution by Theorem 3. Hence, by solving it, we decode the polynomial $a_i(x)$, and by the equality $a_0(x) = S_1(x)+a_i(x)$ we can decode $a_0(x)$ as well. An important observation is that the number of non zero entries in our linear system of equations is exactly $2n-2$, thus the time complexity to solve this linear system of equations is $O (n^2)$~\cite{GostavsonMatrix}.
\end{IEEEproof}

To summarize, given the values of $e_{i,i},e_{0,0}$, an efficient decoding procedure with time complexity $\Theta(n^2)$ works as follows:
\begin{enumerate}
\item Compute $S_1(x),S_2(x)$.
\item Compute $S_1(x)+S_2(x)$.
\item Solve the linear system of equations induced from the equality 
$$S_1(x)+S_2(x)\equiv a_i(x)+a_i(x)x^i (\bmod (x^n-1 ))$$
in order to decode $a_i(x)$. 
\item Use the equality $a_0(x) = S_1(x)+a_i(x)$ in order to decode $a_0(x)$.
\end{enumerate}

Now all that is left to show in order to prove Theorem \ref{Th4} is the decoding of $e_{0,0},e_{i,i}$. This will be done in two steps; first we will decode the values of $e_{i,n-i},e_{0,\ipn{2i}}$ and then we will derive the values of $e_{0,0},e_{i,i}$. The former edges will be decoded using the following algorithm.
%Now all that is left to show in order to prove Theorem \ref{Th4} is the decoding of $e_{0,\ipn{2i}},e_{i,n-i}$. We decode these edges using the following algorithm.
\begin{algorithm}[H]
	\caption{Decoding of $e_{0,\ipn{2i}}$}
	\begin{algorithmic}[1]
		\State Decode $e_{0,i}$ using the $D_i$ constraint
		\State $\ell=3$
		\State $\textrm{sum}=e_{0,i}$
		\While {$\ell < n-1$}
			\State Compute $d_\ell=e_{0,\ipn{\ell\cdot i}} +e_{i,\ipn{\ell\cdot i}}$
			\State Compute $f_\ell=e_{i,\ipn{\ell\cdot i}}+e_{0,\ipn{(\ell+1)\cdot i}}$
			\State $\textrm{sum}=\textrm{sum}+d_\ell+f_\ell$
			\State $\ell=\ell+2$
		\EndWhile
		\State$e_{0,\ipn{2i}}=\textrm{sum}$
\end{algorithmic}\label{alg}
\end{algorithm}

Using a similar algorithm we decode the value $e_{i,n-i}$ as well. To prove the correctness of Algorithm~\ref{alg}, it suffices that we prove the following claim.
\begin{claim}
All steps in Algorithm~\ref{alg} are possible to compute and furthermore, $\textrm{sum}=e_{0,\ipn{2i}}$. 
\end{claim}
\begin{IEEEproof}
First note that the edge $e_{0,i}$ can be decoded according to the $D_i$ diagonal constraint since all the edges in this constraint  are known besides $e_{0,i}$. The values $\ell$ receives in the while loop of the algorithm are $3,5,\ldots,n-2$ and for every value of $\ell$ it is possible to compute $d_\ell$ by the neighborhood constraint of $S_{\ipn{\ell\cdot i}}$. Similarly, the value of $f_\ell$ is computed by the diagonal constraint $D_{\ipn{(\ell+1)\cdot i}}$. 

From the while loop of Algorithm~\ref{alg}, we have that 
\begin{align*}
\textrm{sum}&=e_{0,i}+\sum\limits_{k=1}^{\frac{n-3}{2}} (d_{2k+1}+f_{2k+1}) \\
& =e_{0,i}+\sum\limits_{k=1}^{\frac{n-3}{2}} (e_{0,\ipn{(2k+1)\cdot i}}+e_{0,\ipn{(2k+2)\cdot i}})\\
&=\sum\limits_{\ell=1,\ell\neq 2}^{n-1} e_{0,\ipn{\ell\cdot i}}\overset{(a)}{=}\sum\limits_{\ell=1,\ell\neq \ipn{2i}}^{n-1} e_{0,\ell}\overset{(b)}{=}e_{0,\ipn{2i}}.
\end{align*}
Step (a) holds since $i$ is a generator of the group $\Z _n$, and thus $\set{\ipn{3i},\ipn{4i},\ldots,\ipn{(n-1)\cdot i}}$ are all distinct elements in $\Z _n$, and since we also added the term $e_{0,i}$ to this summation. Lastly, Step (b) holds by the neighborhood constraint of $S_0$ and we get that $\textrm{sum}=e_{0,\ipn{2i}}$. \end{IEEEproof}
We are now read to conclude with the proof of Theorem~\ref{Th4}.

\begin{figure*} 
	\centering
	\subfigure[Slope-Two-Diagonal-Parity Constraints on $A_G$]{\includegraphics[width=80mm]{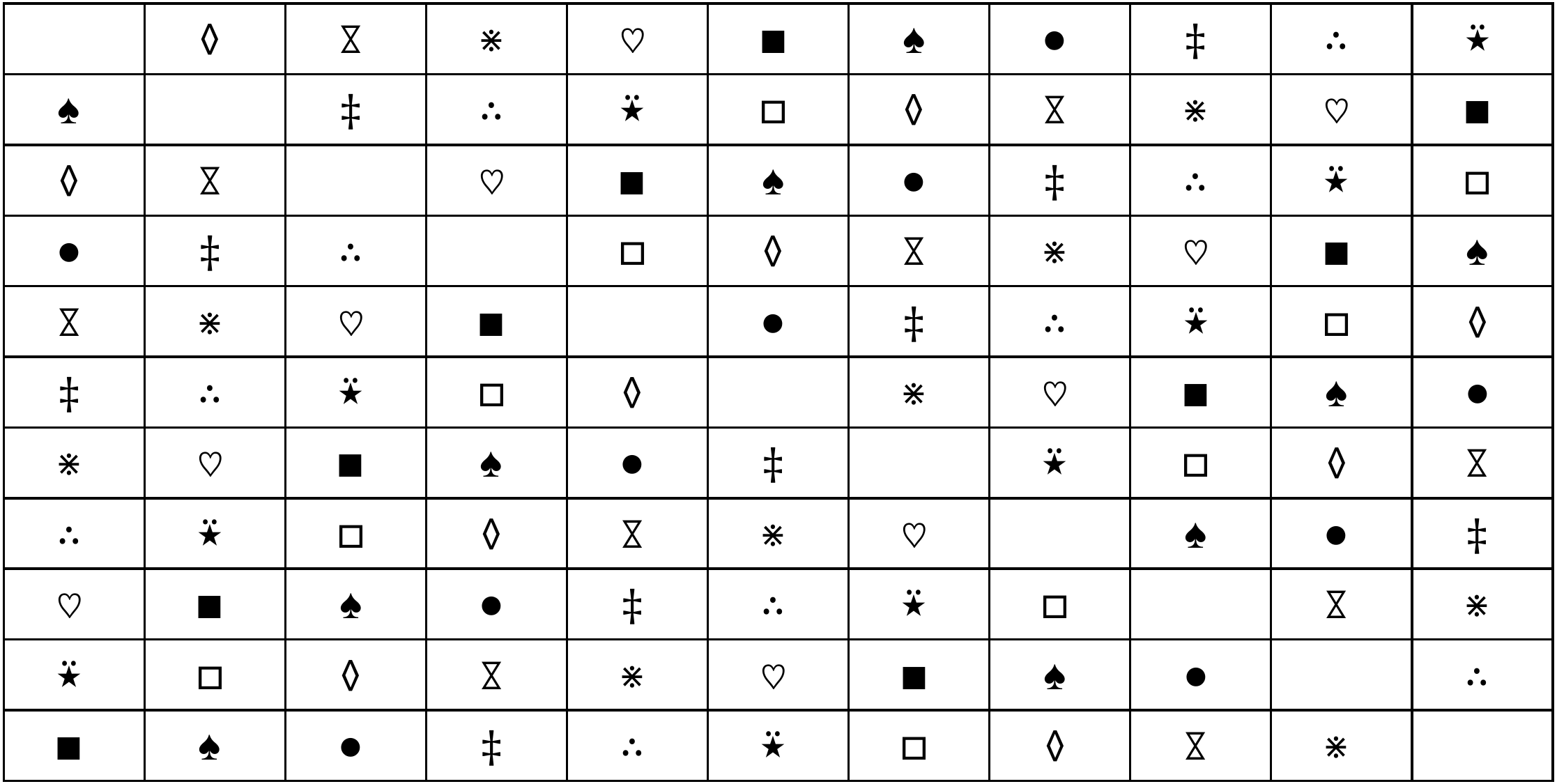}}~~~~~~~~~~~~
	\subfigure[Slope-Two-Diagonal-Parity Constraints on $A'_G$]{\includegraphics[width=80mm]{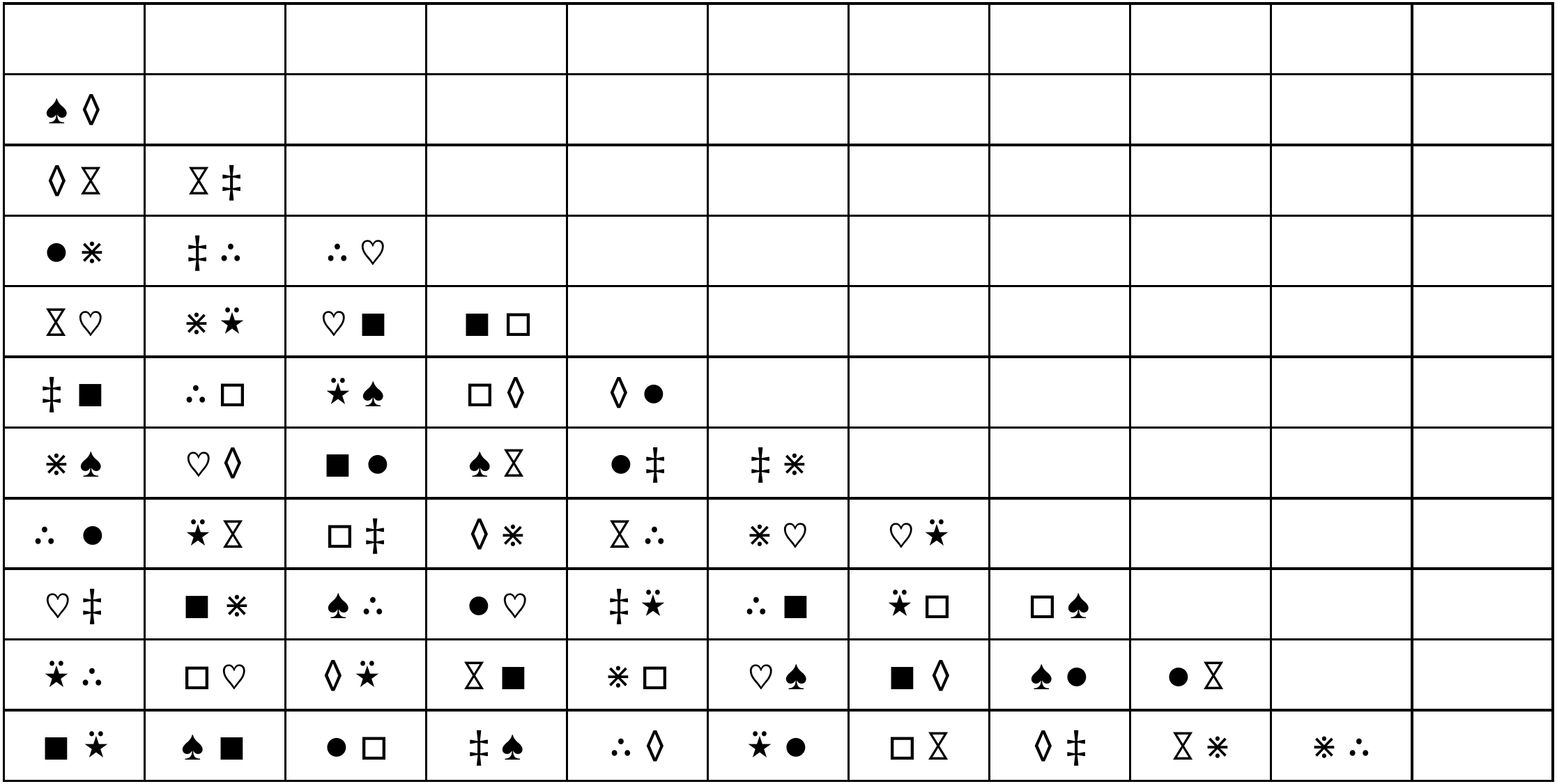}}
	\caption{The slope-two-diagonal constraints over undirected graphs, represented on the labeling matrix and the lower-triangle-labeling matrix.}\label{fig:ex3}
	%The constraints of undirected graph $G$, represented on its labeling matrix $A_G$, and its lower-triangle-labeling matrix $A_G'$.}\label{fig:ex3}
\end{figure*}

\begin{IEEEproof}[Proof of Theorem~\ref{Th4}]
    Using a similar algorithm to Algorithm~\ref{alg} we can decode the edge $e_{i,n-i}$ and using the diagonal constraints $D_0,D_{\ipn{2i}}$ we can lastly decode $e_{0,0},e_{i,i}$, from $e_{i,n-i},e_{0,\ipn{2i}}$, respectively. This concludes the proof of the decoding procedure and of Theorem \ref{Th4}.
\end{IEEEproof}

\section{Binary Triple-Node-Erasure-Correcting Codes}\label{sec:three-node}
In this section we present a construction of binary triple-node-erasure-correcting codes for undirected graphs. 
Let $n\geq 5$ be a prime number such that $2$ is a primitive number in $\Z_n$.  Let $G = (V_n,L)$ be a graph with $n$ vertices. We will use in this construction the edge sets $S_h,D_{m}$ for $h\in [n], m\in[n]$ which were defined in~\eqref{S_h},\eqref{D_m}, respectively. In addition, for $s\in[n]$ we define the edge set
  \begin{equation*}
  T_{s} = \{\langle v_k,v_{\ell} \rangle|k,\ell \in[n],  \langle k + 2\ell \rangle_n = s,k\neq \ell \}.
  \end{equation*}
  In this construction we impose the same constraints from Construction~\ref{const:double}, that is, the sets $S_{h}$ will be used to represent parity constraints on the neighborhood of each node, the sets $D_{m}$ will represent parity constraints on the diagonals with slope one of $A_G$, and furthermore the sets $T_{s}$ will represent parity constraints on the diagonals with slope two of $A_G$. 

\begin{example}	
In Fig.~\ref{fig:ex3} we present the sets $T_s$, $s\in[11]$ of a graph $G = (V_{11}, L)$ on its labeling matrix $A_G$, and its lower-triangle-labeling matrix $A_G'$.
\end{example}

We are now ready to show the following construction.
\begin{Construction}\label{const:triple}
	For all prime number $n\geq 5$ where $2$ is primitive in $\Z_n$, let $\cC_3$ be the following code: 
	\begin{equation*}
	\cC_3= \left\{G = (V_n,L)  \middle|
	\begin{array}{cc}
	(a)& \sum_{\langle v_i,v_j \rangle\in S_h}e_{i,j}= 0, h \in [n]  \\ 
	(b)& \sum_{\langle v_i,v_j \rangle\in D_m}e_{i,j}=0, m\in[n] \\
	(c)& \sum_{\langle v_i,v_j \rangle\in T_s}e_{i,j}=0, s\in[n] 	
	\end{array}
	\right\}.
	\end{equation*}
\end{Construction}
Note that the code $\cC_3$ is a subcode of the code $\cC_2$ and for \emph{any} graph $G$ over the binary field, by~\eqref{double_red} there are only $n-1$ independent constraints (a) in Construction~\ref{const:triple}, and by the same principle,
\begin{align*}
	\sum_{s\in[n]}\sum_{\langle v_i,v_j \rangle\in T_s} e_{i,j}  = \sum_{s=0}^{n-1}\hspace{-2ex}\sum_{ \substack{ \ell=0 \\\ell\neq \langle 3^{-1}s \rangle_n}}^{n-1} \hspace{-2ex}e_{ \langle s-2\ell \rangle_n,\ell}  
	=  2 \sum_{h=0}^{n-1}\sum_{\ell=0}^{h-1}e_{h,\ell}  = 0.
\end{align*}
Therefore the code $\cC_3$ has at most $3n-2$ linearly independent constraints which implies that its redundancy is not greater than $3n-2$. Since we will prove in Theorem~\ref{th:triple} that $\cC_3$ is a triple-node-correcting codes, according to the Singleton bound we get that the code redundancy is at most a single bit away from optimality.
%upper bound on the code redundancy of the code $\cC_3$ is at most $3n-2$, and thus it is 
Our main result in this section is showing the following theorem.
\begin{theorem}\label{th:triple}
For all prime number $n\geq5$ such that $2$ is primitive in $\Z_n$, the code $\cC_3$ is a triple-node-erasure-correcting code. It is at most a single bit away from optimality. 
\end{theorem}
\begin{IEEEproof}
Assume on the contrary that there is a graph $G = (V_n,L) \in \cC_3$ where $w(G) \leq 3$. We prove here only the case that $w(G) = 3$ since the case of $w(G) \leq 2$ holds according to Theorem~\ref{th:double}. By the symmetry of Construction~\ref{const:triple}, it is sufficient to assume that a vertex cover $W$ of $G$ is  $W = \{v_0,v_i,v_j\}$ for distinct $i,j\in[n]\setminus \{0\}$, while all other cases hold by relabeling the indices $0,i,j$. We will show that $G = G_{\textbf{0}}$. 
 
Denote by $H_{i,j} = \{i,j,\langle 2i \rangle_n,\langle 2j \rangle_n,\langle 2i+j \rangle_n,\langle 2j+i \rangle_n\}$. This set represents all the special constraints of $T_\ell$, $\ell \in H_{i,j}$, that will be used in the following claim.
\begin{claim}The following properties hold on the graph $G$:\label{claim:first_prop}
	\begin{enumerate}[(a)]
		\item For all $\ell\in[n]\setminus \{0,i,j \}$, $e_{0,\ell}+e_{i,\ell}+e_{j,\ell} = 0$. \label{claim:first_prop1}
		\item For all $\ell\in[n]\setminus \{i,j,\langle i+j \rangle_n \}$, $e_{0,\ell}+e_{i,\langle \ell-i \rangle_n}+e_{j,\langle \ell-j \rangle_n} = 0$.\label{claim:first_prop2}
		\item $e_{0,i} + e_{j,\langle i-j \rangle_n} = 0, e_{0,j} + e_{i,\langle j-i \rangle_n} = 0$ and $e_{j,i} + e_{0,\langle i+j \rangle_n} = 0$.\label{claim:first_prop3}
		\item For all $\ell\in[n]\setminus H_{i,j}$,\label{claim:first_prop4}
		\begin{align*}
		 & e_{0,\ell}+e_{i,\langle \ell -2i \rangle_n}+e_{j,\langle \ell -2j \rangle_n} \\
		 & +e_{0,\langle 2^{-1}\ell \rangle_n}+e_{i,\langle 2^{-1}(\ell-i) \rangle_n}+e_{j,\langle 2^{-1}(\ell-j) \rangle_n} = 0.
		\end{align*}
		\item\label{claim:first_prop5}
		\begin{align*}
		 &\sum_{\ell \in H_{i,j}}\Big( e_{0,\ell}+e_{i,\langle \ell -2i \rangle_n}+e_{j,\langle \ell -2j \rangle_n} \\
		 & +e_{0,\langle 2^{-1}\ell \rangle_n}+e_{i,\langle 2^{-1}(\ell-i) \rangle_n}+e_{j,\langle 2^{-1}(\ell-j)  \rangle_n}\Big)x^{\ell}  \\
		 &\equiv  e_{i,0}(x^i+x^{2i}) + e_{j,0}(x^j+x^{2j}) + e_{j,i}(x^{2i+j}+x^{i+2j}) \\
		 & ~~~~~~~~~~~~~~~~~~~~~~~~~~~~~~~~~~~~~~~~~~~~~~~~~~~~~~~~~~~~~~~~~~~(\bmod x^n-1).		 	
		\end{align*}
	\end{enumerate}
\end{claim}

The proofs of Claim~\ref{claim:first_prop}\ref{claim:first_prop1},~\ref{claim:first_prop2}, and \ref{claim:first_prop3} are very similar to the proof of Claim~\ref{cl1}, so they are deferred to Appendix~\ref{app:0}.

\begin{IEEEproof} 
		We remind that for all $s,\ell \in[n] \setminus \{0,i,j\}$, $e_{s,\ell} = 0$.
		\begin{enumerate}[(a),start = 4]		
		\item For all $\ell\in[n]$,  let $B_{\ell}$ be the following edge set
		\begin{align}\label{H_l}
		& B_{\ell} = \{ \langle v_{0}, v_{\ell} \rangle, \langle v_{i}, v_{\langle \ell-2i \rangle_n} \rangle, \langle v_{j}, v_{\langle \ell-2j \rangle_n} \rangle,  \\
		&\nonumber \langle v_{0},v_{\langle 2^{-1}\ell \rangle_n} \rangle, \langle v_{i}, v_{\langle 2^{-1}(\ell-i) \rangle_n} \rangle, \langle v_{j}, v_{\langle 2^{-1}(\ell-j) \rangle_n} \rangle   \}.
		\end{align}
		 It can be readily verified that for $\ell \notin \{0,\langle 3i \rangle_n,\langle 3j \rangle_n \}\cup H_{i,j}$, $|B_{\ell}|=6$.  For  all $s\in\{0,i,j\}$ and for all $\ell\in[n]\setminus  \{0,\langle 3i \rangle_n,\langle 3j \rangle_n \} $ it holds that  $s\neq \langle \ell-2s \rangle_n$ and therefore, if $\langle v_{s},v_{\langle \ell-2s \rangle_n}\rangle \in B_{\ell}$ then $\langle v_{s},v_{\langle \ell-2s \rangle_n}\rangle \in T_{\ell}$, i.e., $B_\ell \subseteq T_\ell$. 
		% since $\langle v_{s},v_{\langle \ell-2s \rangle_n}\rangle \in T_{\ell}, s\neq \langle \ell-2s \rangle_n$ and $\langle v_{s},v_{\langle 2^{-1}(\ell-s) \rangle_n }\rangle \in T_{\ell}, s\neq \langle 2^{-1}(\ell-s)  \rangle_n$, we deduce that for  $\ell \notin \{0,\langle 3i \rangle_n,\langle 3j \rangle_n \}$, $H_{\ell} \subseteq T_{\ell}$, and for $\ell \in \{0,\langle 3i \rangle_n,\langle 3j \rangle_n \}$, $H_{\ell}\setminus \{e_{\ell,\ell}\} \subseteq T_{\ell}$ where the edge set $H_{\ell}\setminus \{e_{\ell,\ell}\}$ consists of all the edges incident  to at least one of the nodes $v_{0},v_{i}$ and $v_{j}$ in $T_{\ell}$.
		  Therefore, by the definition of the diagonal constraint $(c)$ in Construction~\ref{const:triple} we deduce that for all $\ell \notin \{0,\langle 3i \rangle_n,\langle 3j \rangle_n \}\cup H_{i,j}$,
		\begin{align*}
		0	 = &\sum_{\langle v_k, v_m \rangle \in T_\ell}e_{k,m } =\sum_{\langle v_k, v_m \rangle \in B_\ell}e_{k,m } \\
		 = & e_{0,\ell}+e_{i,\langle \ell -2i \rangle_n}+e_{j,\langle \ell -2j \rangle_n} \\
		 + & e_{0,\langle 2^{-1}\ell \rangle_n}+e_{i,\langle 2^{-1}(\ell-i) \rangle_n}+e_{j,\langle 2^{-1}(\ell-j) \rangle_n}.
		\end{align*}
		 Moreover, for $\ell = 0$, $\langle v_{0}, v_{\ell} \rangle =\langle v_{0},v_{\langle 2^{-1}\ell \rangle_n} \rangle = \langle v_{0}, v_{0} \rangle$ and therefore $|B_{0}|=5$. It can be similarly verified that $|B_{\langle 3i \rangle_n}| =|B_{\langle 3j \rangle_n}| = 5 $. Notice that  for all $s\in\{0,i,j\}, \ell \in \{0,\langle 3i \rangle_n,\langle 3j \rangle_n \}$, if $\langle v_{s},v_{\langle \ell-2s \rangle_n}\rangle \in B_{\ell}$ then it holds that $\langle v_{s},v_{\langle \ell-2s \rangle_n}\rangle \in T_{\ell}\cup \{\langle v_s,v_s \rangle \}$, i.e., $B_\ell \subseteq T_\ell\cup \{\langle v_s,v_s \rangle \}$.
		 Therefore again, by the definition of the diagonal constraint $(c)$ in Construction~\ref{const:triple} we deduce that for all  $\ell \in \{0,\langle 3i \rangle_n,\langle 3j \rangle_n \}$,
		 \begin{align*}
		 0 = &\sum_{\langle v_k, v_m \rangle \in T_\ell\cup \{\langle v_{\langle 3^{-1}\ell \rangle_n},v_{\langle 3^{-1}\ell \rangle_n}\rangle \}}e_{k,m } +e_{\langle 3^{-1}\ell \rangle_n,\langle 3^{-1}\ell \rangle_n} \\
		  =&\sum_{\langle v_k, v_m \rangle \in B_\ell}e_{k,m }+e_{\langle 3^{-1}\ell \rangle_n,\langle 3^{-1}\ell \rangle_n} \\
		 =&  e_{0,\ell}+e_{i,\langle \ell -2i \rangle_n}+e_{j,\langle \ell -2j \rangle_n} \\
		 +& e_{0,\langle 2^{-1}\ell \rangle_n}+e_{i,\langle 2^{-1}(\ell-i) \rangle_n}+e_{j,\langle 2^{-1}(\ell-j) \rangle_n} .
		 \end{align*}
		\item For all $\ell \in H_{i,j}$ let $B_\ell$ be the edge set from~\eqref{H_l}. Notice that for $\ell = i$ we get that $\langle v_{0}, v_{\ell} \rangle = \langle v_{i}, v_{\langle 2^{-1}(\ell-i) \rangle_n} \rangle$, for $\ell = \langle 2i \rangle_n$ we get that $\langle v_{i}, v_{\langle \ell-2i \rangle_n} \rangle= \langle v_{0},v_{\langle 2^{-1}\ell \rangle_n} \rangle$, and for $\ell = \langle 2i+j \rangle_n$ we get that $\langle v_{i}, v_{\langle \ell-2i \rangle_n} \rangle=  \langle v_{j}, v_{\langle 2^{-1}(\ell-j) \rangle_n} \rangle $, and therefore for all $\ell \in  H_{i,j}$, $|B_{\ell}|=5$. Similarly to the proof of $(d)$, the edge set $B_{\ell}$ consists of all the edges incident  to at least one of the nodes $v_{0},v_{i}$ and $v_{j}$ in $T_{\ell}$, i.e., $B_\ell \subseteq T_\ell$. Therefore we deduce that for $\ell \in \{i,j\}$, 
		 \begin{align*}
		 e_{\ell,0} = &\sum_{\langle v_k, v_m \rangle \in T_\ell}e_{k,m } +e_{\ell,0} \\
		 =&\sum_{\langle v_k, v_m \rangle \in B_\ell}e_{k,m }+e_{\ell,0} \\
		 =&  e_{0,\ell}+e_{i,\langle \ell -2i \rangle_n}+e_{j,\langle \ell -2j \rangle_n} \\
		 +& e_{0,\langle 2^{-1}\ell \rangle_n}+e_{i,\langle 2^{-1}(\ell-i) \rangle_n}+e_{j,\langle 2^{-1}(\ell-j) \rangle_n},
		 \end{align*}
		and the coefficient of the monomial $x^i, x^j$ in the polynomial 
		\begin{align*}
		&\sum_{\ell \in H_{i,j}}\Big( e_{0,\ell}+e_{i,\langle \ell -2i \rangle_n}+e_{j,\langle \ell -2j \rangle_n} \\
		& +e_{0,\langle 2^{-1}\ell \rangle_n}+e_{i,\langle 2^{-1}(\ell-i) \rangle_n}+e_{j,\langle 2^{-1}(\ell-j)  \rangle_n}\Big)x^{\ell} \\
		 & ~~~~~~~~~~~~~~~~~~~~~~~~~~~~~~~~~~~~~~~~~~~~~~~~~~(\bmod x^n-1)	
		\end{align*}
		is $e_{i,0},e_{j,0}$, respectively. The proof that the coefficient of $x^{2i}, x^{2j},x^{2i+j}, x^{2j+i}$ in this polynomial is $e_{i,0},e_{j,0},e_{j,i},e_{j,i}$ is similar, respectively.
		\end{enumerate}	
\end{IEEEproof}	

Let  $a_0(x), a_i(x)$ and $ a_j(x)$ be the neighborhood polynomials without self-loops of $G$. We are ready to construct the following equation system.
\begin{lemma}\label{claim:poly} The following properties hold:
	\begin{enumerate}[(a)]
		\item $ a_0(x) + a_i(x) + a_j(x)   $\label{eq:sys6_2}\\ 
		$= e_{i,0}(1+x^i) + e_{j,0}(1+x^j) +  e_{j,i}(x^i+x^j) $.
		\item $a_0(x) + a_i(x)x^i + a_j(x)x^j $ \label{eq:sys6_3} \\
		$   \equiv e_{0,0}+e_{i,i}x^{2i} + e_{j,j}x^{2j} +  e_{i,0}x^i +  e_{j,0}x^j +  e_{j,i}x^{i+j} \\~~~~~~~~~~~~~~~~~~~~~~~~~~~~~~~~~~~~~~~~~~~~~~~~~~~~~~~~~~~~~~~~~~~~~~~~~~~~~~~~~~ (\bmod x^n-1)$.
		\item $a_0(x) + a_i(x)x^{2i} + a_j(x)x^{2j}  + a^2_0(x) + a^2_i(x)x^i + a^2_j(x)x^j $ \label{eq:sys6_4} \\
		 $\equiv  e_{i,0}(x^i+x^{2i}) + e_{j,0}(x^j+x^{2j}) + e_{j,i}(x^{2i+j}+x^{i+2j})
		 \\~~~~~~~~~~~~~~~~~~~~~~~~~~~~~~~~~~~~~~~~~~~~~~~~~~~~~~~~~~~~~~~~~~~~~~~~~~~~~~~~~~     (\bmod x^n-1)$.
	\end{enumerate}
%\begin{align} 
%\nonumber & = e_{j,0} + e_{k,0} + (e_{j,0} + e_{k,j})x^j + (e_{k,0} + e_{k,j})x^k (\bmod x^n-1)\\
%\nonumber & = e_{j,0}(1+x^j) + e_{k,0}(1+x^k) +  e_{k,j}(x^j+x^k)  (\bmod x^n-1),\\
%\nonumber & = e_{0,0}+e_{j,j}x^{2j} + e_{k,k}x^{2k} +  e_{j,0}x^j +  e_{k,0}x^k +  e_{k,j}x^{j+k}(\bmod x^n-1),\\ 
%\nonumber & =  e_{j,0}x^j + e_{k,0}x^k + e_{k,j}x^{2j+k} + e_{j,0}x^{2j} + e_{k,0}x^{2k} + e_{k,j}x^{j+2k}   (\bmod x^n-1) \\
%\nonumber & =  e_{j,0}(x^j+x^{2j}) + e_{k,0}(x^k+x^{2k}) + e_{k,j}(x^{2j+k}+x^{j+2k})    (\bmod x^n-1).
%\end{align}	
\end{lemma}
\begin{IEEEproof}
	\begin{enumerate}[(a)]
	\item 
	\begin{align*}
	& a_0(x) + a_i(x) + a_j(x) =  \\
	& = e_{0,0}+e_{i,i}x^i + e_{j,j}x^j + \sum^{n-1}_{\ell = 0}e_{0,\ell}x^{\ell}+\sum^{n-1}_{\ell = 0}e_{i,\ell}x^{\ell}+\sum^{n-1}_{\ell = 0}e_{j,\ell}x^{\ell}\\
	 & = e_{0,0}+e_{i,i}x^i + e_{j,j}x^j + \sum^{n-1}_{\ell = 0}\Big(e_{0,\ell}+e_{i,\ell}+e_{j,\ell} \Big)x^{\ell}\\
	& \stackrel{(a)}= e_{0,0}+e_{i,i}x^i + e_{j,j}x^j   + \Big(e_{0,0}+e_{i,0}+e_{j,0} \Big) \\
	& + \Big(e_{0,i}+e_{i,i}+e_{j,i} \Big)x^i + \Big(e_{0,j}+e_{i,j}+e_{j,j} \Big)x^j  \\
	&= e_{i,0}(1+x^i) + e_{j,0}(1+x^j) +  e_{j,i}(x^i+x^j),
	\end{align*}
	where Step~(a) holds since by Claim~\ref{claim:first_prop}\ref{claim:first_prop1} for all $\ell\in[n]\setminus \{0,i,j \}$ the coefficient of $x^{\ell}$ is zero.

	\item %By definition,
	\begin{align*}
	& a_0(x) + a_i(x)x^i + a_j(x)x^j = \\
	& = e_{0,0}+e_{i,i}x^{2i} + e_{j,j}x^{2j}  \\
	&+ \sum^{n-1}_{\ell = 0}e_{0,\ell}x^{\ell}+\sum^{n-1}_{\ell = 0}e_{i,\ell}x^{\ell+i}+\sum^{n-1}_{\ell = 0}e_{j,\ell}x^{\ell+j}\\
	&\equiv e_{0,0}+e_{i,i}x^{2i} + e_{j,j}x^{2j}  \\
	&+ \sum^{n-1}_{\ell = 0}e_{0,\ell}x^{\ell}+\sum^{n-1}_{\ell = 0}e_{i,\langle \ell -i \rangle_n}x^{\ell}+\sum^{n-1}_{\ell = 0}e_{j,\langle \ell -j \rangle_n}x^{\ell} (\bmod x^n-1)\\
	&\equiv e_{0,0}+e_{i,i}x^{2i} + e_{j,j}x^{2j}  \\
	&+ \sum^{n-1}_{\ell = 0}\Big(e_{0,\ell}+e_{i,\langle \ell -i \rangle_n}+e_{j,\langle \ell -j \rangle_n}\Big)x^{\ell}(\bmod x^n-1) \\
	&\stackrel{(a)}  \equiv e_{0,0}+e_{i,i}x^{2i} + e_{j,j}x^{2j}  \\
	&+ \Big(e_{0,i}+e_{i,0}+e_{j,\langle i -j \rangle_n}\Big)x^i + \Big(e_{0,j}+e_{i,\langle j -i \rangle_n }+e_{j,0}\Big)x^j\\
	&+ \Big(e_{0,\langle i+j\rangle_n}+e_{i,j}+e_{j,i}\Big)x^{i+j} (\bmod x^n-1) \\
	& \stackrel{(b)} \equiv e_{0,0}+e_{i,i}x^{2i} + e_{j,j}x^{2j}  \\
	&+ e_{i,0}x^i + e_{j,0 }x^j+ e_{j,i}x^{i+j} (\bmod x^n-1).
	\end{align*}
	Step~(a) holds since  by Claim~\ref{claim:first_prop}\ref{claim:first_prop2} for all $\ell\in[n]\setminus \{i,j,\langle i+j \rangle_n \}$ the coefficient of $x^{\ell}$ is zero, and Step~(b)
	holds since by  Claim~\ref{claim:first_prop}\ref{claim:first_prop3} we know that $e_{0,i} = e_{j,\langle i-j \rangle_n},e_{0,j} = e_{i,\langle j-i \rangle_n}$, and $e_{j,i} = e_{0,\langle i+j \rangle_n}$.

	\item According to the neighborhood-polynomials definition we can write 
	\begin{align*}
	& a_0(x) + a_i(x)x^{2i} + a_j(x)x^{2j}  + a^2_0(x) + a^2_i(x)x^i + a^2_j(x)x^j  \\
	&= e_{0,0}+e_{i,i}x^{3i} + e_{j,j}x^{3j}  \\
	&+ \sum^{n-1}_{\ell = 0}e_{0,\ell}x^{\ell}+\sum^{n-1}_{\ell = 0}e_{i,\ell}x^{\ell+2i}+\sum^{n-1}_{\ell = 0}e_{j,\ell}x^{\ell+2j}\\
	&+ e_{0,0}+e_{i,i}x^{3i} + e_{j,j}x^{3j}  \\
	&+ \sum^{n-1}_{\ell = 0}e_{0,\ell}x^{2\ell}+\sum^{n-1}_{\ell = 0}e_{i,\ell}x^{2\ell+i}+\sum^{n-1}_{\ell = 0}e_{j,\ell}x^{2\ell+j} \\
	&\equiv \sum^{n-1}_{\ell = 0}e_{0,\ell}x^{\ell}+\sum^{n-1}_{\ell = 0}e_{i,\langle \ell -2i \rangle_n}x^{\ell}+\sum^{n-1}_{\ell = 0}e_{j,\langle \ell -2j \rangle_n}x^{\ell}\\
	&+ \sum^{n-1}_{\ell = 0}e_{0,\langle 2^{-1}\ell \rangle_n}x^{\ell}+\sum^{n-1}_{\ell = 0}e_{i,\langle 2^{-1}(\ell-i) \rangle_n}x^{\ell}+\sum^{n-1}_{\ell = 0}e_{j,\langle 2^{-1}(\ell-j) \rangle_n}x^{\ell}
\\	&~~~~~~~~~~~~~~~~~~~~~~~~~~~~~~~~~~~~~~~~~~~~~~~~~~~(\bmod x^n-1)\\
	&\equiv \sum^{n-1}_{\ell = 0}\Big( e_{0,\ell}+e_{i,\langle \ell -2i \rangle_n}+e_{j,\langle \ell -2j \rangle_n} \\
	& +e_{0,\langle 2^{-1}\ell \rangle_n}+e_{i,\langle 2^{-1}(\ell-i) \rangle_n}+e_{j,\langle 2^{-1}(\ell-j) \rangle_n}\Big)x^{\ell} \\
	 &~~~~~~~~~~~~~~~~~~~~~~~~~~~~~~~~~~~~~~~~~~~~~~~~~~~(\bmod x^n-1)\\
	&\stackrel{(a)}\equiv \sum_{\ell \in H_{i,j}}\Big( e_{0,\ell}+e_{i,\langle \ell -2i \rangle_n}+e_{j,\langle \ell -2j \rangle_n} \\
	& +e_{0,\langle 2^{-1}\ell \rangle_n}+e_{i,\langle 2^{-1}(\ell-i) \rangle_n}+e_{j,\langle 2^{-1}(\ell-j)  \rangle_n}\Big)x^{\ell} \\
	&\stackrel{(b)}\equiv  e_{i,0}(x^i+x^{2i}) + e_{j,0}(x^j+x^{2j}) + e_{j,i}(x^{2i+j}+x^{i+2j})  \\  
	&~~~~~~~~~~~~~~~~~~~~~~~~~~~~~~~~~~~~~~~~~~~~~~~~~~~(\bmod x^n-1),			 	
	\end{align*}
	where Step~(a) holds since by Claim~\ref{claim:first_prop}\ref{claim:first_prop4} for all $\ell\in[n]\setminus H_{i,j}$ the coefficient of $x^{\ell}$ is zero, and Step~(b) is a direct result of Claim~\ref{claim:first_prop}\ref{claim:first_prop5}.
	\end{enumerate}	
\end{IEEEproof}

Notice that by setting $x=1$ in the equation of Lemma~\ref{claim:poly}\ref{eq:sys6_3} we get that
\begin{align}\label{eq:allsum}
 e_{0,0}+ e_{i,i}+ e_{j,j}+ e_{i,0}+ e_{j,0}+ e_{j,i} = 0.
\end{align}

Using the result of Lemma~\ref{claim:poly} we get the next three equalities. The proof of this lemma is given in Appendix~\ref{app:1}
\begin{lemma} \label{claim:eq2} The following equations hold
	\begin{enumerate}[(a)]
		\item $ a_j(x)(1+x^i) +  a^2_j(x)\equiv e_{j,j}(1+x^{j})(x^{i}+x^{j})\\    
		~~~~~~~~~~~~~~~~~~~~~~~~~~~~~~~~~~~~~~~~~~~~~~~~~~~~~~~~~~~~~~~~~~  (\bmod x^n-1)$.\label{claim:eq2_1} 
		\item $ a_i(x)(1+x^j) +  a^2_i(x) \equiv e_{i,i}(1+x^{i})(x^{i}+x^{j})\\    
		~~~~~~~~~~~~~~~~~~~~~~~~~~~~~~~~~~~~~~~~~~~~~~~~~~~~~~~~~~~~~~~~~~    (\bmod x^n-1)$.\label{claim:eq2_2} 
		\item $ a_0(x)(x^i+x^j) +  a^2_0(x) \equiv e_{0,0}(1+x^{i})(1+x^{j}) \\    
		~~~~~~~~~~~~~~~~~~~~~~~~~~~~~~~~~~~~~~~~~~~~~~~~~~~~~~~~~~~~~~~~~~   (\bmod x^n-1)$.\label{claim:eq2_3} 
	\end{enumerate}
\end{lemma}

Our next step is showing that the value of at least one of the self-loops $e_{j,j},e_{i,i}$ or $e_{0,0}$ is zero. For this goal, we show another important claim where its proof is given in Appendix~\ref{app:3}.
\begin{lemma}\label{cor:imp}
It holds that  $e_{0,0} + e_{i,i} + e_{j,j} = e_{j,0} + e_{j,0} + e_{j,i} = 0$. 
%	The value of one of the self-loops $e_{j,j}, e_{i,i}$ or $e_{0,0}$ is zero, and the value of one of the edges $e_{j,i},e_{i,0}$ and $e_{j,0}$ is zero.
\end{lemma}	

By Lemma~\ref{cor:imp}, we know that at least one of the self-loops $e_{j,j}, e_{i,i}$ or $e_{0,0}$ is zero, and our next step is showing that one of the polynomials $a_0(x), a_i(x)$ or $a_j(x)$ is zero. We assume that $e_{j,j}$ is zero, while the proof of the other two cases will be similar based upon Lemma~\ref{claim:eq2}\ref{claim:eq2_2} and~\ref{claim:eq2}\ref{claim:eq2_3}. By Lemma~\ref{claim:eq2}\ref{claim:eq2_1}, we get that
\begin{align*}
 a_j(x)[1+x^i +  a_j(x)] \equiv 0  (\bmod x^n-1).
\end{align*}
Denote by $p(x)$ the polynomial $p(x) = 1+x^i +  a_j(x)$ which is clearly in $\cR_n$. Since $M_n(x)$ is irreducible, either $M_n(x)|a_j(x)$ or $M_n(x)|p(x)$. Since $1+x|a_j(x)$ and $1+x|p(x)$ it is possible to derive that either $a_j(x) = 0 $ or $p(x) = 0$. We will show that $p(x) \neq 0$ which will lead to the fact that $a_j(x) = 0$. 
Assume on a contrary that $p(x) = 0$. Therefore we deduce that $ a_j(x) = 1+x^i $ and thus $e_{j,i} = e_{j,0} = 1$. 
Notice that in this case, by Lemma~\ref{cor:imp} we have that $e_{i,0}=0$. 
By Lemma~\ref{claim:poly}\ref{eq:sys6_2} we deduce that
\begin{align*}
		& a_0(x) + a_i(x) + 1+x^i     \\ 
		& = a_0(x) + a_i(x) + a_j(x)   \\ 
		& = e_{i,0}(1+x^i) + e_{j,0}(1+x^j) +  e_{j,i}(x^i+x^j) \\
		& = (1+x^j) +  (x^i+x^j) = 1+x^i,		 
\end{align*}
and therefore $a_0(x) + a_i(x) = 0$. Again, by Lemma~\ref{cor:imp} we know that  $e_{0,0}+ e_{i,i} + e_{j,j} = 0$ and therefore, since $e_{j,j}=0$, we get that $e_{i,i} = e_{0,0}$. By Lemma~\ref{claim:poly}\ref{eq:sys6_3} we deduce that
\begin{align*}
	& a_0(x) + a_i(x)x^i + (1+x^i)x^j  \\	
	& = a_0(x) + a_i(x)x^i + a_j(x)x^j   \\
	&\equiv e_{0,0}+e_{i,i}x^{2i} + e_{j,j}x^{2j} +  e_{i,0}x^i +  e_{j,0}x^j +  e_{j,i}x^{i+j} (\bmod x^n-1)   \\
 	& \equiv e_{0,0}+e_{i,i}x^{2i}   +  x^j +  x^{i+j}  (\bmod x^n-1)  \\
 	& \equiv e_{0,0}+e_{i,i}x^{2i}   +  (1+x^i)x^j  (\bmod x^n-1)  ,
\end{align*}
and therefore $a_0(x) + a_i(x)x^i \equiv e_{0,0}+e_{i,i}x^{2i} (\bmod x^n-1)$. As we showed in the proof of Theorem~\ref{th:double}, since the conditions of Claim~\ref{cl2} hold, we deduce also here that 
%   for the correctness of Claim~\ref{cl2} in the proof of Theorem~\ref{th:double}}, in this case we have that 
$a_0(x)=a_i(x)=0$, and therefore we get a contradiction since $e_{j,i} = e_{j,0} = 1$. Therefore, it holds that  $a_j(x) = 0$ and since $\cC_3$ is a sub code of $\cC_2$, we again get that $a_0(x)= a_i(x)=0$, and that cocludes the proof.  
\end{IEEEproof}

\begin{comment}
\section{Rebuilding Ratio of a Single Erasure}
Assume that only $v_0$ has failed, we would like to compute exactly how many edges we would need to read in order to restore $v_0$. We analyze this by decoding some of the edges in $N(v_0)$ using the neighborhood constraints, and the rest by the diagonal constraints. Assume we decode the $x$ last edges in $N(v_0)$ with the neighborhood constraints and the rest by the diagonal constraints(by the natural order on the edges).  
\begin{claim}
	We read exactly $(n-1-x)\cdot x +\frac{x^2-x}{2}$ edges when restoring the last $x$ edges of $v_0$ using the neighborhood constraints
\end{claim}
\emph{Proof.} Observe the matrix representation of the graph, we can see that in order to restore the last $x$ edges of $v_0$ using the neighborhood constraints we need to read the rectangle with length and width $n-1-x$ and $x$ respectively, and an additional 90 degree triangle with 2 edges of length $x$. Clearly the amount of edges in the rectangle is $(n-1-x)\cdot x$. All that is left to prove is that the amount of edges in the triangle is exactly $\frac{x^2-x}{2}$.

\end{comment}

\section{Conclusion}\label{sec:conc}
In this paper we continued our research on codes over graphs from~\cite{YY17,YY18}. We presented an optimal binary construction for codes correcting a failure of two nodes together with a decoding procedure that is complexity optimal.  We then extended this construction for triple-node-erasure-correcting codes which are at most a single bit away from optimality with respect to the Singleton bound.

\section*{Acknowledgments}
The authors would like to thank Gil Kupfer for his contribution to the result of Corollary~\ref{lem:hard} in Appendix~\ref{app:3}.

\appendices
\section{}\label{app:0}
In this appendix we prove the first three properties of Claim~\ref{claim:first_prop}.
\begin{customclaim}{\ref{claim:first_prop}}The following properties hold on the graph $G$:
	\begin{enumerate}[(a)]
		\item For all $\ell\in[n]\setminus \{0,i,j \}$, $e_{0,\ell}+e_{i,\ell}+e_{j,\ell} = 0$. 
		\item For all $\ell\in[n]\setminus \{i,j,\langle i+j \rangle_n \}$, $e_{0,\ell}+e_{i,\langle \ell-i \rangle_n}+e_{j,\langle \ell-j \rangle_n} = 0$.
		\item $e_{0,i} + e_{j,\langle i-j \rangle_n} = 0, e_{0,j} + e_{i,\langle j-i \rangle_n} = 0$ and $e_{j,i} + e_{0,\langle i+j \rangle_n} = 0$.
		\item For all $\ell\in[n]\setminus H_{i,j}$,
		\begin{align*}
		 & e_{0,\ell}+e_{i,\langle \ell -2i \rangle_n}+e_{j,\langle \ell -2j \rangle_n} \\
		 & +e_{0,\langle 2^{-1}\ell \rangle_n}+e_{i,\langle 2^{-1}(\ell-i) \rangle_n}+e_{j,\langle 2^{-1}(\ell-j) \rangle_n} = 0.
		\end{align*}
		\item
		\begin{align*}
		 &\sum_{\ell \in H_{i,j}}\Big( e_{0,\ell}+e_{i,\langle \ell -2i \rangle_n}+e_{j,\langle \ell -2j \rangle_n} \\
		 & +e_{0,\langle 2^{-1}\ell \rangle_n}+e_{i,\langle 2^{-1}(\ell-i) \rangle_n}+e_{j,\langle 2^{-1}(\ell-j)  \rangle_n}\Big)x^{\ell}  \\
		 &\equiv  e_{i,0}(x^i+x^{2i}) + e_{j,0}(x^j+x^{2j}) + e_{j,i}(x^{2i+j}+x^{i+2j}) \\
		 & ~~~~~~~~~~~~~~~~~~~~~~~~~~~~~~~~~~~~~~~~~~~~~~~~~(\bmod x^n-1).			 	
		\end{align*}
		\end{enumerate}
\end{customclaim}
\begin{IEEEproof} 
	We remind that for all $s,\ell \in[n] \setminus \{0,i,j\}$, $e_{s,\ell} = 0$.
	\begin{enumerate}[(a)]		
		\item We know that for all  $\ell\in[n]\setminus \{0,i,j \}, s\in[n]\setminus \{\ell \}$, $\langle v_{s},v_{\ell}\rangle \in S_{\ell}$, and therefore by the definition of the constraint $(a)$ in Construction~\ref{const:triple} we get that
		\begin{align*}
		&0 =\sum_{\langle v_s,v_\ell \rangle\in S_{\ell} }e_{s,\ell}=\sum^{n-1}_{s=0,s\neq\ell}e_{s,\ell} = e_{0,\ell}+e_{i,\ell}+e_{j,\ell}.
		\end{align*}
		\item For all $m\in [n]\setminus \{i,j,i+j\}$, denote  by $D'_m$ the set
			$$D'_m = D_m\backslash \set{ \langle v_0,v_m \rangle,\langle v_i,v_{\langle m-i \rangle_n }\rangle ,\langle v_j,v_{\langle m-j \rangle_n }\rangle }. $$
			Therefore, we have that 
			\begin{align*}
			&0 = \sum\limits _{\langle  v_j,v_{\langle m-j \rangle_n} \rangle \in D_m} e_{j,\langle m-j \rangle_n} = \\
			&\sum\limits _{\langle  v_j,v_{\langle m-j \rangle_n} \rangle \in D'_m} e_{j,\langle m-j \rangle_n}   +e_{0,m}+e_{i,\langle m-i \rangle_n}+e_{j,\langle m-j \rangle_n},
			\end{align*}
		and since $e_{s,\ell} =0$ for all $\langle v_s,v_{\ell} \rangle \in  D'_m$, we get that $e_{0,m} +e_{i,\langle m-i\rangle_n}+e_{i,\langle m-i\rangle_n}=0$.
		\item Similarly to $(b)$, for $m = i$ we get that $\langle v_0,v_{m} \rangle  = \langle v_{i},v_{\langle m-i \rangle_n} \rangle$ and therefore by the definition of the constraint $(b)$ in Construction~\ref{const:triple} we get that  $e_{0,i}+e_{j,\langle i-j \rangle_n} = 0$.
		It can be similarly verified that for $m = j$ we get that $e_{0,j} + e_{i,\langle j-i \rangle_n} = 0$ and for $m = \langle i+j \rangle_n$ we get that $e_{j,i} + e_{0,\langle i+j \rangle_n} = 0$.
	\end{enumerate}
\end{IEEEproof}

\section{}\label{app:1}
 Remember that $i$ and $j$ are indices such that $i,j\in[n]\setminus\{0\}$.
\begin{customlemma} {\ref{claim:eq2}} The following equations hold
	\begin{enumerate}[(a)]
		\item $ a_j(x)(1+x^i) +  a^2_j(x)\equiv e_{j,j}(1+x^{j})(x^{i}+x^{j})\\    
		~~~~~~~~~~~~~~~~~~~~~~~~~~~~~~~~~~~~~~~~~~~~~~~~~~~~~~~~~~~~~~~~~~  (\bmod x^n-1)$.
		\item $ a_i(x)(1+x^j) +  a^2_i(x) \equiv e_{i,i}(1+x^{i})(x^{i}+x^{j})\\    
		~~~~~~~~~~~~~~~~~~~~~~~~~~~~~~~~~~~~~~~~~~~~~~~~~~~~~~~~~~~~~~~~~~    (\bmod x^n-1)$.
		\item $ a_0(x)(x^i+x^j) +  a^2_0(x) \equiv e_{0,0}(1+x^{i})(1+x^{j}) \\    
		~~~~~~~~~~~~~~~~~~~~~~~~~~~~~~~~~~~~~~~~~~~~~~~~~~~~~~~~~~~~~~~~~~   (\bmod x^n-1)$. 
	\end{enumerate}
\end{customlemma}

\begin{IEEEproof}
We only prove equation~\ref{claim:eq2_1} while the other two hold by relabelling of the construction. 
	 First we prove two useful properties on polynomials.
	\begin{claim}\label{claim:algeb2}
		The following equation holds
		\begin{align*}
		&(x^{2i}+x^{2j})(1+x^i) + (1+x^{2j})(1+x^i)x^{2i}  \\
		& +(x^{2i}+x^{2j})^2 + (1+x^{2j})^2x^i  \\
		&=(1+x^i)(1+x^{2j})(x^{2i}+x^{2j})  +(1+x^i)^{3}x^i.
		\end{align*}
	\end{claim}
	
	\begin{IEEEproof}
		This equation can be rewritten by 
		\begin{align*}
		&(x^{2i}+x^{2j})(1+x^i) + (1+x^{2j})(1+x^i)x^{2i} \\
		& + (1+x^i)(1+x^{2j})(x^{2i}+x^{2j})  +(1+x^i)^{3}x^i \\
		&= (x^{2i}+x^{2j})^2 + (1+x^{2j})^2x^i,
		\end{align*}
		or,
		\begin{align*}
		&(1+x^i) [(x^{2i}+x^{2j}) + (1+x^{2j})x^{2i} + \\
		&+(1+x^{2j})(x^{2i}+x^{2j})  +(1+x^i)^{2}x^i] \\
		&= (x^{2i}+x^{2j})^2 + (1+x^{2j})^2x^i.
		\end{align*}
		Moreover, it can be rewritten by
		\begin{align*}
		&(1+x^i) [x^{2i}+x^{2j} + x^{2i}+x^{2j+2i} + \\
		&+(1+x^{2j})(x^{2i}+x^{2j})  +x^i+x^{3i}] \\
		&= (x^{2i}+x^{2j})^2 + (1+x^{2j})^2x^i,
		\end{align*}
		or
		\begin{align*}
		&(1+x^i) [x^{2j} +x^{2j+2i} + \\
		&+x^{2i}+x^{2j}+x^{2i+2j}+x^{4j} +x^i+x^{3i}] \\
		&= x^{4i}+x^{4j} + x^i+x^{4j+i}.
		\end{align*}
		We finally rewrite it by 
		\begin{align*}
		&(1+x^i) (x^{2i}+x^{4j} +x^i+x^{3i})  \\
		&= x^{4i}+x^{4j} + x^i+x^{4j+i},
		\end{align*}
		which holds since $(1+x^i) (x^{2i}+x^{4j} +x^i+x^{3i}) = x^{2i}+x^{4j} +x^i+x^{3i} + x^{3i}+x^{4j+i} +x^{2i}+x^{4i} =  x^{4i}+x^{4j} + x^i+x^{4j+i} $.
	\end{IEEEproof}		
	
	Let $e_{0,0},e_{i,i},e_{j,j}$ be the label on the self-loop of node $v_0,v_i,v_j$, respectively, and let $e_{i,0},e_{j,0},e_{j,i}$ be the label on the edge $\langle v_i,v_0 \rangle,\langle v_j,v_0 \rangle,\langle v_j,v_i \rangle$, respectively. We define the following three polynomials
	\begin{align*}
	& p(x) =   e_{i,i}(1+x^{2i}) + e_{j,j}(1+x^{2j}) + e_{j,i}(1+x^i)(1+x^j), \\
	& q(x) =  e_{0,0}(1+x^{2i}) + e_{j,j}(x^{2i}+x^{2j}) +  e_{j,0}(x^i+x^j)(1+x^i), \\
	& s(x) = e_{i,0}(x^i+x^{2i}) + e_{j,0}(x^j+x^{2j}) + e_{j,i}(x^{2i+j}+x^{i+2j}).
	\end{align*}

	\begin{claim}\label{claim:algeb}
		The following equation holds
		\begin{align*}
		& q(x)(1+x^i) + p(x)(1+x^i)x^{2i} + q^2(x) + p^2(x)x^i\\
		& +s(x)(1+x^i)^2 =  e_{j,j}(1+x^i)(1+x^{2j})(x^{2i}+x^{2j}).
		\end{align*}
	\end{claim}
	
	\begin{IEEEproof}
		\begin{align*}
		& q(x)(1+x^i) + p(x)(1+x^i)x^{2i} + q^2(x) + p^2(x)x^i\\
		& +s(x)(1+x^i)^2 = \\
		& [e_{0,0}(1+x^{2i}) + e_{j,j}(x^{2i}+x^{2j}) +  e_{j,0}(x^i+x^j)(1+x^i)](1+x^i) \\
		& +[e_{i,i}(1+x^{2i}) + e_{j,j}(1+x^{2j}) + e_{j,i}(1+x^i)(1+x^j)](1+x^i)x^{2i} \\
		& +[ e_{0,0}(1+x^{2i}) + e_{j,j}(x^{2i}+x^{2j}) +  e_{j,0}(x^i+x^j)(1+x^i)]^2 \\
		& +[e_{i,i}(1+x^{2i}) + e_{j,j}(1+x^{2j}) + e_{j,i}(1+x^i)(1+x^j)]^2x^i \\
		& +[e_{i,0}(x^i+x^{2i}) + e_{j,0}(x^j+x^{2j}) + e_{j,i}(x^{2i+j}+x^{i+2j})](1+x^i)^2 \\
		& = e_{0,0}(1+x^i)^{3}x^i + e_{i,i}[(1+x^i)^{3}x^{2i} + (1+x)^{4i}x^{i}]  \\
		& + e_{i,0}(x^i+x^{2i})(1+x^i)^2 \\
		& +  e_{j,0}(1+x^i)^2[(x^i+x^j)+(x^i+x^j)^2+(x^j+x^{2j})] \\
		& + e_{j,i}(1+x^i)^2[(1+x^j)x^{2i}+(1+x^j)^2x^{i}+(x^{2i+j}+x^{i+2j})] \\
		&  + e_{j,j}[(x^{2i}+x^{2j})(1+x^i) + (1+x^{2j})(1+x^i)x^{2i}  \\
		& ~~~~~~~~~~~~~~~~~~~~~~~~~~~~~~~~~~+ (x^{2i}+x^{2j})^2 + (1+x^{2j})^2x^i] \\
		&\stackrel{(a)} = e_{0,0}(1+x^i)^{3}x^i + e_{i,i}(1+x^i)^{3}x^i  +  e_{i,0}(1+x^i)^{3}x^i \\
		& + e_{j,0}(1+x^i)^2(x^i+x^{2i}) \\
		& + e_{j,i}(1+x^i)^2[x^{2i} + x^{2i+j}+x^{i}+x^{i+2j}+x^{2i+j}+x^{i+2j}] \\
		& +  e_{j,j}[(1+x^i)(1+x^{2j})(x^{2i}+x^{2j})  +(1+x^i)^{3}x^i] \\
		& = e_{0,0}(1+x^i)^{3}x^i + e_{i,i}(1+x^i)^{3}x^i  +  e_{i,0}(1+x^i)^{3}x^i \\
		& +  e_{j,0}(1+x^i)^{3}x^i + e_{j,i}(1+x^i)^{3}x^i \\
		& +  e_{j,j}[(1+x^i)(1+x^{2j})(x^{2i}+x^{2j})  +(1+x^i)^{3}x^i] \\
		&\stackrel{(b)} =  e_{j,j}(1+x^i)(1+x^{2j})(x^{2i}+x^{2j}),
		\end{align*}
		where Step~(a) holds since by Claim~\ref{claim:algeb2},
		\begin{align*}
		& e_{j,j}[(x^{2i}+x^{2j})(1+x^i) + (1+x^{2j})(1+x^i)x^{2i} + \\
		& (x^{2i}+x^{2j})^2 + (1+x^{2j})^2x^i] \\
		& = e_{j,j}[(1+x^i)(1+x^{2j})(x^{2i}+x^{2j})  +(1+x^i)^{3}x^i],
		\end{align*}
		and Step~(b) holds since by equation~\eqref{eq:allsum}  $ e_{0,0} + e_{i,i}+ e_{j,j}+ e_{i,0}+ e_{j,0}+ e_{j,i} = 0 $.
	\end{IEEEproof}
	
	Summing the equation of~Lemma~\ref{claim:poly}\ref{eq:sys6_2} with the equation of~Lemma~\ref{claim:poly}\ref{eq:sys6_3} we get
	\begin{align*} 
	&  a_i(x)(1+x^i) + a_j(x)(1+x^j) \equiv e_{0,0}+e_{i,i}x^{2i} + e_{j,j}x^{2j}   \\
	&\nonumber   +e_{i,0} +  e_{j,0} +  e_{j,i}(x^i+x^j+x^{i+j})  (\bmod x^n-1),
	\end{align*}
	and since $ e_{0,0} = e_{i,i}+ e_{j,j}+ e_{i,0}+ e_{j,0}+ e_{j,i} $ we rewrite it as
	\begin{align} \label{eq:sys7_1}
	&  a_i(x)(1+x^i) + a_j(x)(1+x^j) \equiv e_{i,i}(1+x^{2i})    \\
	&   \nonumber + e_{j,j}(1+x^{2j}) + e_{j,i}(1+x^i)(1+x^j) = p(x) (\bmod x^n-1).
	\end{align}
	Multiplying the equation of Lemma~\ref{claim:poly}\ref{eq:sys6_2} by $x^i$ and adding it to the equation of Lemma~\ref{claim:poly}\ref{eq:sys6_3}  we get
	\begin{align*}
	&  a_0(x)(1+x^i) + a_j(x)(x^i+x^j) \equiv e_{0,0}+e_{i,i}x^{2i} + e_{j,j}x^{2j} \\
	&\nonumber  + e_{i,0}x^{2i} +  e_{j,0}(x^i+x^j+x^{i+j}) +  e_{j,i}x^{2i}  (\bmod x^n-1),
	\end{align*}
	and since $ e_{i,i}  =e_{0,0}+ e_{j,j}+ e_{i,0}+ e_{j,0}+ e_{j,i} $ we rewrite it as
	\begin{align}\label{eq:sys7_2}
	&  a_0(x)(1+x^i) + a_j(x)(x^i+x^j)\equiv e_{0,0}(1+x^{2i}) \\
	&\nonumber   + e_{j,j}(x^{2i}+x^{2j}) +  e_{j,0}(x^i+x^j)(1+x^i) =q(x)   (\bmod x^n-1).
	\end{align}
	Next, we multiply the equation of Lemma~\ref{claim:poly}\ref{eq:sys6_4} by $(x^i+1)^2$. In the left-hand side of this equation we set the value of $a_0(x)(1+x^i)$ from equation~\eqref{eq:sys7_2} and the value of $ a_i(x)(1+x^i)$ from equation~\eqref{eq:sys7_1} to get that %in the left hand side of this equation to  and we show its left hand side by
	\begin{align*}
	& a_0(x)(1+x^i)^2 + a_i(x)(1+x^i)^2x^{2i} + a_j(x)(1+x^i)^2x^{2j} \\
	& + a^2_0(x)(1+x^i)^2 + a^2_i(x)(1+x^i)^2x^i + a^2_j(x)(1+x^i)^2x^j \\
	& \equiv [a_j(x)(x^i+x^j) + q(x) ](1+x^i) \\
	& +[a_j(x)(1+x^j) + p(x)  ](1+x^i)x^{2i}  \\
	& +[a_j(x)(x^i+x^j) + q(x) ]^2 \\
	& +[a_j(x)(1+x^j) + p(x) ]^2 x^i \\
	&+ a_j(x)(1+x^i)^2x^{2j}  + a^2_j(x)(1+x^i)^2x^j (\bmod x^n-1).
	\end{align*}
	The right-hand side of this equation is
	\begin{align*}
	s(x)(1+x^i)^2 (\bmod x^n-1).
	\end{align*}
	Now, we proceed with the calculations, while having on the left-hand side only the values that depend on $a_j(x)$, so we receive that,
	\begin{align*}
	&  a_j(x)[ (x^i+x^j)(1+x^i) +(1+x^j)(1+x^i)x^{2i}+(1+x^i)^2x^{2j} ] \\
	& + a^2_j(x)[(x^i+x^j)^2+(1+x^j)^2x^i+(1+x^i)^2x^j]\\
	& \equiv  a_j(x)(1+x^i)(x^i+x^j+x^{2i}+x^{2j}+x^{2i+j}+x^{i+2j}) \\
	& + a^2_j(x)(x^i+x^j+x^{2i}+x^{2j}+x^{2i+j}+x^{i+2j}) (\bmod x^n-1)\\
	& \equiv  a_j(x)(1+x^i)^2(1+x^j)(x^i+x^j)  \\
	& +  a^2_j(x)(1+x^i)(1+x^j)(x^i+x^j)  (\bmod x^n-1).
	\end{align*}  
	The right-hand side of the last equation is rewritten to be 
	\begin{align*}
	& q(x)(1+x^i) + p(x)(1+x^i)x^{2i} + q^2(x) + p^2(x)x^i\\
	& +s(x)(1+x^i)^2 (\bmod x^n-1),
	\end{align*}
	which is equal to $e_{j,j}(1+x^i)(1+x^{2j})(x^{2i}+x^{2j})$ by Claim~\ref{claim:algeb}. Combining both sides together we deduce that
	\begin{align*}
	&  a_j(x)(1+x^i)^2(1+x^j)(x^i+x^j) \\
	&+  a^2_j(x)(1+x^i)(1+x^j)(x^i+x^j) \\
	&\nonumber\equiv e_{j,j}(1+x^i)(1+x^{2j})(x^{2i}+x^{2j})  (\bmod x^n-1),
	\end{align*}
	which can be rewritten by,
	\begin{align*}
	&  (1+x^i)(1+x^{j})(x^{i}+x^{j})\\
	&  \cdot[a_j(x)(1+x^i) +  a^2_j(x) + e_{j,j}(1+x^{j})(x^{i}+x^{j})]\equiv 0\\
	&   ~~~~~~~~~~~~~~~~~~~~~~~~~~~~~~~~~~~~~~~~~~~~~~~~~~~~ (\bmod x^n-1).
	\end{align*}
	Lastly, denote by $m(x)$ the polynomial $$m(x) = a_j(x)(1+x^i) +  a^2_j(x) + e_{j,j}(1+x^{j})(x^{i}+x^{j}),$$ where it holds that $1+x|m(x)$ since $m(1)\equiv 0 (\bmod x^n-1)$. Notice that the polynomials $1+x^i$ and $1+x^j$ are in $\cR_n$ and by~\eqref{Mn_prop1} they are co-prime to $M_n(x)$. Similarly, the polynomial $x^i+x^j$ is also in $\cR_n$ and thus is co-prime to $M_n(x)$ as well. Therefore, we deduce that $M_n(x)|m(x)$ and $m(x)\equiv 0 (\bmod x^n-1)$, which leads to,
	\begin{align*}
	&  a_j(x)(1+x^i) +  a^2_j(x) \equiv e_{j,j}(1+x^{j})(x^{i}+x^{j})(\bmod x^n-1). 
	% = e_{j,j}(x^{i}+x^{j}+x^{i+j}+x^{2j}) 
	\end{align*}
	
\end{IEEEproof}

\section{}\label{app:3}
\begin{customlemma}{\ref{cor:imp}}
It holds that  $e_{0,0} + e_{i,i} + e_{j,j} = e_{j,0} + e_{j,0} + e_{j,i} = 0$. 
%	The value of one of the self-loops $e_{j,j}, e_{i,i}$ or $e_{0,0}$ is zero, and the value of one of the edges $e_{j,i},e_{i,0}$ and $e_{j,0}$ is zero.
\end{customlemma}	
\begin{IEEEproof}
	We start with proving several important claims.
	\begin{claim}\label{claim:eq33} If 
		\begin{align}\label{claim:eq3}
		&  a_j(x)(1+x^i) +  a^2_j(x)   \\
		&\nonumber \equiv e_{j,j}(x^{j}+x^{i+j}+x^{2j})  (\bmod x^n-1),
		\end{align}
		then for all $s\in [n]$
		\begin{align}\label{eq:eq4}
		e_{j,s}=e_{j,\langle 2s \rangle_n}+e_{j,\langle 2s-i \rangle_n},
		\end{align}
		and for all $1\leq t \leq n-1 $ it holds that
		\begin{align}\label{eq:eq11115}
		e_{j,s}  = \sum^{2^t-1}_{\ell=0}e_{j,\langle 2^ts-\ell i \rangle_n} .
		\end{align}
	\end{claim}
	\begin{IEEEproof}
		First notice that by calculating the coefficient of $x^{\langle 2s \rangle_n}$ of equation \eqref{claim:eq3} for all $s\in [n]$ such that $\langle 2s \rangle_n\notin \{j,\langle i+j \rangle_n,\langle 2j \rangle_n \}$ it holds that 
		\begin{align*}
		e_{j,\langle 2s \rangle_n}+e_{j,\langle 2s-i \rangle_n} + e_{j,s} = 0.
		\end{align*}
		For $\langle 2s \rangle_n = j,\langle 2s \rangle_n = \langle i+j \rangle_n,\langle 2s \rangle_n = \langle 2j \rangle_n$, since the coefficient of $x^j,x^{\langle i+j \rangle_n},x^{\langle 2j \rangle_n}$ in $a_j(x),a_j(x)x^i,a^2_j(x)$ is zero, respectively, we deduce that also in this case we can write
		\begin{align*}
		e_{j,\langle 2s \rangle_n}+e_{j,\langle 2s-i \rangle_n} + e_{j,s} = 0,
		\end{align*}
		which proves the correctness of equation~\eqref{eq:eq4}.
		
		Next, we prove the rest of this claim by induction on $t$ where $1\leq t \leq n-1$.\\
		\textbf{Base:} for $t=1$, as we showed above, by calculating the coefficient of $x^{\langle 2s \rangle_n}$ of equation \eqref{claim:eq3} we deduce that for all $s\in[n]$ it holds 
		\begin{align*}
		e_{j,s}=e_{j,\langle 2s \rangle_n}+e_{j,\langle 2s-i \rangle_n}.
		\end{align*}
		\textbf{Step:} assume that the claim holds for all $\tau$ where $1\leq \tau \leq t-1\leq n-2$, that is, 	
		\begin{align*}
		e_{j,s}  =  \sum^{2^{\tau}-1}_{\ell=0}e_{j,\langle 2^{\tau}s-\ell i \rangle_n}.
		\end{align*}
		By the correctness of equation~\eqref{eq:eq4} and replacing $s$ with $\langle 2^{\tau}s-\ell i \rangle_n$ we deduce that 	
		$$e_{j,\langle 2^{\tau}s-\ell i \rangle_n}=e_{j,\langle 2(2^{\tau}s-\ell i) \rangle_n}+e_{j,\langle 2(2^{\tau}s-\ell i) - i \rangle_n},$$ 
		and for $\tau=t-1$ we get that
		\begin{align*}
		&	e_{j,s} = \sum^{2^{t-1}-1}_{\ell=0}e_{j,\langle 2^{t-1}s-\ell i \rangle_n}  \\
		& 	 = \sum^{2^{t-1}-1}_{\ell=0}\Big( e_{j,\langle 2^ts-2\ell i \rangle_n} + e_{j,\langle 2^ts-2\ell i- i  \rangle_n}\Big) =   \sum^{2^{t}-1}_{\ell=0}e_{j,\langle 2^{t}s-\ell i \rangle_n}.
		\end{align*}
	\end{IEEEproof}
	For $a,t,s\in[n]$ denote by  $I_{a,t,s}$ the number
	$$ I_{a,t,s} = |\{(\tau,\ell)~|~ \langle 2^{\tau}s-\ell a \rangle_n = \langle   a2^{-1} \rangle_n, \tau\in[t], \ell \in [2^{\tau}]\}|.$$
	%Notice that for all  $a_1,b_1,t\in[n]$ and $a_2,b_2,t\in[n]$ it holds that $I_{a_1,b_1,t,0} = I_{a_2,b_2,t,0}$.
	
	\begin{corollary}\label{cor:cor1}
		For all $1\leq t \leq n-1 $ and $s\in[n]$ it holds that
		\begin{align*}
		e_{j,s}  = \sum^{2^t-1}_{\ell=0}e_{j,\langle 2^ts-\ell i \rangle_n}  + e_{j,j}I_{i,t,s}.
		\end{align*}
	\end{corollary}
	
	\begin{IEEEproof}
		By adding the monomial $e_{j,j}x^i$ to equation~\eqref{claim:eq3} we get the same expression as in Claim~\ref{claim:eq2}\ref{claim:eq2_1} on the right-hand side.  According to equation~\eqref{eq:eq4}, by calculating the coefficient of $x^{\langle 2s \rangle_n}$,  for all $s\in[n]\setminus \{\langle 2^{-1}i \rangle_n \}$ we will get $	e_{j,s}  =  e_{j,\langle 2s \rangle_n}+e_{j,\langle 2s-i \rangle_n}$ and for $s = \langle 2^{-1}i \rangle_n$ we will get  $	e_{j,s}  =  e_{j,\langle 2s \rangle_n}+e_{j,\langle 2s-i \rangle_n}+ e_{j,j}$. According to the modification of equation~\eqref{eq:eq4} for $s = \langle 2^{-1}i \rangle_n$, we need to similarly adjust equation~\eqref{eq:eq11115} by counting the number of times the self-loop $e_{j,j}$ should be added to the equation. Hence, by the same arguments of the proof of Claim~\ref{claim:eq33}, we deduce that 
		\begin{align*}
		e_{j,s}  = \sum^{2^t-1}_{\ell=0}e_{j,\langle 2^ts-\ell i \rangle_n} + e_{j,j}I_{i,t,s},
		\end{align*}
		where by definition, $I_{i,t,s}$ is the number of pairs $(\tau,\ell)$ where $\tau \in[t]$ and $\ell \in[2^{\tau}]$ such that $\langle 2^{\tau}s-\ell i \rangle_n = \langle i2^{-1} \rangle_n$.
	\end{IEEEproof}
	
	Next we show another important claim.
	%	e_{i,js} + e_{j,is}  = 	e_{i,j-js} + e_{j,i-is}  + e_{i,i} + e_{j,j}.
	\begin{claim}\label{claim:eq5}
		For all $s \in [n]$ it holds
		\begin{align*}
		e_{j,s}  = e_{j,\langle i-s \rangle_n} + e_{j,j}(1+I_{i,\frac{n-1}{2},s}).
		\end{align*}

	\end{claim}
	\begin{IEEEproof}
		By Corollary~\ref{cor:cor1} we know that for all $t,s\in[n]$,
		\begin{align*}
		e_{j,s}  = \sum^{2^t-1}_{\ell=0}e_{j,\langle 2^ts-\ell i \rangle_n}  + e_{j,j}I_{i,t,s}.
		\end{align*}
		Since $2$ is primitive in $\Z_n$, there exists a $t$ for which ${\langle 2^{t}\rangle_n=n-1}$, or equivalently, there is an odd positive number $h$ such that $2^{t}=hn-1$. It can be verified that in this case $t=(n-1)/2$. Therefore,
		\begin{align*}
		& e_{j,s}  = \sum^{nh-2}_{\ell=0}e_{j,\langle -s-\ell i \rangle_n}  + e_{j,j}I_{i,\frac{n-1}{2},s} \\
		&\nonumber = \sum^{n(h-1)-1}_{\ell=0}e_{j,\langle -s-\ell i \rangle_n}+\sum^{nh-2}_{\ell=nh-n}e_{j,\langle -s-\ell i \rangle_n}  + e_{j,j}I_{i,\frac{n-1}{2},s} \\
		&\nonumber = \sum^{n(h-1)-1}_{\ell=0}e_{j,\langle -s-\ell i \rangle_n}+\sum^{n-2}_{\ell=0}e_{j,\langle -s-\ell i \rangle_n}  + e_{j,j}I_{i,\frac{n-1}{2},s} \\
		&\nonumber \stackrel{(a)} = \sum^{n-2}_{\ell=0}e_{j,\langle -s-\ell i \rangle_n}  + e_{j,j}I_{i,\frac{n-1}{2},s} \\
		&\nonumber  = e_{j,\langle i-s \rangle_n}  +\sum^{n-1}_{\ell=0}e_{j,\langle -s-\ell i \rangle_n}+ e_{j,j}I_{i,\frac{n-1}{2},s} \\
		&\nonumber \stackrel{(b)}= e_{j,\langle i-s \rangle_n} + e_{j,j}+ e_{j,j}I_{i,\frac{n-1}{2},s} \\
		&\nonumber= e_{j,\langle i-s \rangle_n} + e_{j,j}(1+I_{i,\frac{n-1}{2},s}).
		\end{align*}
		Note that the summation $\sum^{n(h-1)-1}_{\ell=0}e_{j,\langle -s-\ell i \rangle_n}$ expresses the neighborhood of the $j$th node (including its self-loop) $h-1$ times (i.e., an even number of times). Hence, in Step (a) we noticed that $\sum^{n(h-1)-1}_{\ell=0}e_{j,\langle -s-\ell i \rangle_n} = 0$. Step $(b)$ holds since  $\sum^{n-1}_{\ell=0}e_{j,\ell} = e_{j,j}$.
	\end{IEEEproof}	
	
	Our next goal is to show that the value of $I_{i,\frac{n-1}{2},i}$ is even. For that we show two more claims. First, we define for $t\in[\frac{n+1}{2}]$ the indicator bit $x_t$ as follows: 
	\[x_t = \begin{cases} 
		0 & \textmd{if}~\langle 2^{t-1} \rangle_n < \langle 2^{-1} \rangle_n, \\
		1 & \textmd{if}~\langle 2^{t-1} \rangle_n \geq \langle 2^{-1} \rangle_n.
	\end{cases}	
	\]

	\begin{claim}\label{claim:even2}
		For all $2 \leq t \leq \frac{n-1}{2} $,
			\begin{align*}
		I_{i,t,i} - I_{i,t - 1,i} =2(I_{i,t-1,i} - I_{i,t - 2,i}) - x_{t-1} + x_t.
		\end{align*}
	\end{claim}
	\begin{IEEEproof}
		By definition, for $t \in [\frac{n-1}{2}]$, $I_{i,t,i}$ is given by
		\begin{align*}
		& I_{i,t,i} = |\{(\tau,\ell)| \langle 2^{\tau}i-\ell i \rangle_n = \langle   i2^{-1} \rangle_n, \tau\in[t], \ell \in [2^{\tau}]\}| \\
		&=  |\{(\tau,\ell)| \langle 2^{\tau}-\ell  \rangle_n = \langle   2^{-1} \rangle_n, \tau\in[t], \ell \in [2^{\tau}]\}| \\
		&=  |\{(\tau,m)| \langle m  \rangle_n = \langle   2^{-1} \rangle_n, \tau\in[t], 1\leq m \leq 2^{\tau}\}|.
		\end{align*}
		Therefore, it holds that for  all $2 \leq t \leq \frac{n-1}{2} $
		\begin{align*}
		I_{i,t,i} - I_{i,t - 1,i} 
		& = |\{m~|~ \langle m  \rangle_n = \langle   2^{-1} \rangle_n,   1\leq m \leq 2^{t-1}\}|  \\
		& = \left\lfloor\frac{2^{t-1}}{n}\right\rfloor + x_t = \left\lfloor 2\cdot\Big(\frac{2^{t-2}}{n}\Big)\right\rfloor + x_t \\
		&  = 2\left\lfloor \frac{2^{t-2}}{n}\right\rfloor+ x_{t-1} + x_t \\
		&  \stackrel{(a)} = 2(I_{i,t-1,i} - I_{i,t - 2,i}) - x_{t-1} + x_t,
		\end{align*}
		where in Step (a) we used the property that $I_{i,t,i} - I_{i,t - 1,i}  = \left\lfloor\frac{2^{t-2}}{n}\right\rfloor + x_{t-1} $.
\begin{comment}
		Now we consider the following cases:\textcolor{red}{
		\begin{enumerate}
		\item $\langle 2^{t-2} \rangle_n < \langle 2^{-1} \rangle_n $ and $\langle 2^{t-1} \rangle_n < \langle 2^{-1} \rangle_n $: we get that 
		$$ 	I_{i,t,i} - I_{i,t - 1,i} =2(I_{i,t-1,i} - I_{i,t - 2,i}),$$
		 $\langle 2^{t} \rangle_n < \langle 2^{-1} \rangle_n $ then
		and
		\end{enumerate}
		 if $\langle 2^{t} \rangle_n \geq \langle 2^{-1} \rangle_n $ then
		$$ 	I_{i,t,i} - I_{i,t - 1,i} =2(I_{i,t-1,i} - I_{i,t - 2,i}) + 1,$$
		which is equivalent to equation~\eqref{claim:even2_1}. Similarly, for $\langle 2^{t-1} \rangle_n \geq \langle 2^{-1} \rangle_n $ if $\langle 2^{t} \rangle_n \geq \langle 2^{-1} \rangle_n $ then
		$$ 	I_{i,t,i} - I_{i,t - 1,i} =2(I_{i,t-1,i} - I_{i,t - 2,i}),$$ 
		and if $\langle 2^{t} \rangle_n < \langle 2^{-1} \rangle_n $ then
		$$ 	I_{i,t,i} - I_{i,t - 1,i} =2(I_{i,t-1,i} - I_{i,t - 2,i})-1,$$
		which is equivalent to equation~\eqref{claim:even2_2}.}
\end{comment}
	\end{IEEEproof}

	\begin{claim}\label{claim:even}
		For all $t\in [\frac{n+1}{2}]$ it holds that $\langle I_{i,t,i}+x_t \rangle_2= 0$. % is even if and only if $x_t=0$. % $\langle 2^{\tau} \rangle_n < \langle 2^{-1} \rangle_n $.
	\end{claim}
	\begin{IEEEproof}
		We will prove this claim by  induction on $t \in [\frac{n-1}{2}]$.\\		
		\textbf{Base:} For $t = 0$,  $\langle 2^0 \rangle_n = 1$ which is smaller than $\langle 2^{-1} \rangle_n$ and indeed $I_{i,0,i} +x_0 = 0$. Similarly, for $t = 1$,  $\langle 2^1 \rangle_n = 2$ which is smaller than $\langle 2^{-1} \rangle_n$ for all prime $n \geq5 $ and therefore again $I_{i,1,i} + x_1 = 0$ is even.\\		
		\textbf{Step:} Assume that the claim holds for $t-1$ where $2 \leq t <\frac{n-1}{2}$. % all $\tau'< \tau\in [\frac{n-1}{2}] $.
		By the induction assumption, we have that $\langle I_{i,t-1,i} + x_{t-1} \rangle_2=0$, and by Claim~\ref{claim:even2} we know that 
		$$I_{i,t,i} - I_{i,t - 1,i} =2(I_{i,t-1,i} - I_{i,t - 2,i}) - x_{t-1} + x_t,$$
		or similarly 
		$$ I_{i,t,i} =3I_{i,t-1,i} - 2I_{i,t - 2,i}  - x_{t-1} + x_t,$$
		and therefore
		\begin{align*}
		\langle I_{i,t,i}+x_t \rangle_2 & = \langle 3I_{i,t-1,i} + 2I_{i,t - 2,i}  + x_{t-1} \rangle_2 \\
		&  = \langle I_{i,t-1,i}  + x_{t-1} \rangle_2 = 0.
		\end{align*}
%		 the value of $I_{i,t,i}$ is even if and only if $x_t=0$.
		\begin{comment}
		 Similarly, by the induction assumption, the value of $I_{i,t-1,i}$ is odd if and only if  $x_{t-1}=1$, and by Claim~\ref{claim:even2} we know that in this case
		$$	I_{i,\tau,i} - I_{i,\tau - 1,i} =2(I_{i,\tau-1,i} - I_{i,\tau - 2,i})  -  \Big\lfloor \frac{\langle 2^{-1} \rangle_n} {\langle 2^{\tau} \rangle_n} \Big\rfloor, $$ 
		or similarly 
		$$ I_{i,\tau,i} =3I_{i,\tau-1,i} - 2I_{i,\tau - 2,i}  -  \Big\lfloor \frac{\langle 2^{-1} \rangle_n} {\langle 2^{\tau} \rangle_n} \Big\rfloor,$$
		and therefore again we get that the value of $I_{i,\tau,i}$ is even if and only if $\langle 2^{\tau} \rangle_n < \langle 2^{-1} \rangle_n $.
			\end{comment}
	\end{IEEEproof}

	\begin{corollary}\label{lem:hard}
		The value of $I_{i,\frac{n-1}{2},i}$ is even.
	\end{corollary}
	\begin{IEEEproof}
		By Claim~\ref{claim:even}, it holds that $x_{\frac{n-1}{2}}=0$ since $\langle 2^{\frac{n-1}{2}-1} \rangle_n = \frac{n-1}{2} < \frac{n+1}{2} = \langle 2^{-1} \rangle_n$ and we immediately deduce that $I_{i,\frac{n-1}{2},i} $ is even.
	\end{IEEEproof}
	
	By Claim~\ref{claim:eq5} we know that
	$$  e_{j,s}  = e_{j,\langle i-s \rangle_n} + e_{j,j}(1+I_{i,\frac{n-1}{2},s}).$$
	Since $I_{i,\frac{n-1}{2},i}$ is even, we get
	\begin{align*}
	e_{j,i}  = e_{j,0} + e_{j,j}.
	\end{align*}
	By symmetry of the construction, we also get
	\begin{align*}
	& e_{j,i}  + e_{i,0} + e_{i,i} = 0, \\
	& e_{j,0}  + e_{i,0} + e_{0,0} = 0.
	\end{align*}
	The summation of the last three equalities results with 
	\begin{align*}
	e_{j,j}  + e_{i,i} + e_{0,0} = 0,
	\end{align*}
	and since
	$$ e_{j,j}  + e_{i,i} + e_{0,0} + e_{j,i}  + e_{i,0} + e_{j,0} = 0,$$
	we deduce that 
	$$ e_{j,i}  + e_{i,0} + e_{j,0} = 0.$$
%	Finally, since all the edges of the code are over $\F_2$ we deduce that the value of one of the self-loops $e_{j,j}, e_{i,i}$ or $e_{0,0}$ is zero and the value of one of the edges $e_{j,i},e_{i,0}$ or $e_{j,0}$ is zero.
	
\end{IEEEproof}

\end{document}